\documentclass[11pt,a4paper]{article}

\pdfoutput=1

\usepackage{jheppub}
\usepackage[normalem]{ulem}
\usepackage{slashed}

\def\beq{\begin{equation}}
\def\eeq{\end{equation}}
\def\bea{\begin{eqnarray}}
\def\eea{\end{eqnarray}}
\def\nn{\nonumber}
\def\nl{\nonumber\\}

\def\roughly#1{\mathrel{\raise.3ex\hbox
{$#1$\kern-.75em\lower1ex\hbox{$\sim$}}}}

\def\s{\sqrt{2}}

\def\bctaunu{b \to c \tau^- {\bar\nu}}

\def \cB{{\cal B}}

\def \cM {{\cal M}}
\def \cA {{\cal A}}

\def \cL {{\cal L}}
\def \cH {{\cal H}}
\def \cN {{\cal N}}
\def \tL {\tilde{\cL}}
\def \ep{\epsilon}
\def \s{\sqrt{2}}
\def \de{\delta}

\def \al{\alpha}
\def \be{\beta}
\def \ga{\gamma}
\def \Ga{\Gamma}

\def \la{\lambda}
\def \si{\sigma}
\def \cL{{\cal L}}
\def \cH{{\cal H}}
\def \lp{\left|}
\def \rp{\right|}

\def \ld{\left.}
\def \rd{\right.}
\def \({\left(}
\def \){\right)}
\def \[{\left[}
\def \]{\right]}

\def \ld{\left.}
\def \rd{\right.}
\def \Re{{\rm Re}}
\def \Im{{\rm Im}}
\def \nuta{\nu_\tau}
\def \nutab{\bar{\nu}_\tau}
\def \vp{{\vec p}}
\def \sp{\slashed{p}}

\def \mA0T{\mathcal{A}^{D^*}_{0, T}}
\def \mDs{m_{D^*}}

\allowdisplaybreaks

\begin{document}

\title{\boldmath A Measurable Angular Distribution for \\ ${\bar B} \to
  D^{*} \tau^- {\bar\nu}_\tau$ Decays}

\author[a]{Bhubanjyoti Bhattacharya,}
\author[b]{Alakabha Datta,}
\author[b]{Saeed Kamali}
\author[c]{and David London}
\affiliation[a]{Department of Natural Sciences, Lawrence Technological University, Southfield, MI 48075, USA}
\affiliation[b]{Department of Physics and Astronomy, \\
108 Lewis Hall, University of Mississippi, Oxford, MS 38677-1848, USA, }
\affiliation[c]{Physique des Particules, Universit\'e de Montr\'eal, \\
C.P. 6128, succ.\ centre-ville, Montr\'eal, QC, Canada H3C 3J7}
\emailAdd{bbhattach@ltu.edu}
\emailAdd{datta@phy.olemiss.edu}
\emailAdd{skamali@go.olemiss.edu}
\emailAdd{london@lps.umontreal.ca}

\abstract{At present, the measurements of $R_{D^{(*)}}$ and
  $R_{J/\psi}$ hint at new physics (NP) in $b \to c \tau^- {\bar\nu}$
  decays. The angular distribution of ${\bar B} \to D^* (\to D \pi) \,
  \tau^{-}{\bar\nu}_\tau$ would be useful for getting information about
  the NP, but it cannot be measured. The reason is that the three-momentum
  ${\vec p}_\tau$ cannot be determined precisely since the decay products
  of the $\tau^-$ include an undetected $\nu_\tau$. In this paper, we
  construct a measurable angular distribution by considering the
  additional decay $\tau^- \to \pi^- \nu_\tau$. The full process
  is ${\bar B} \to D^* (\to D \pi') \, \tau^{-} (\to \pi^- \nu_\tau)
  {\bar\nu}_\tau$, which includes three final-state particles whose
  three-momenta can be measured: $D$, $\pi'$, $\pi^-$. The magnitudes
  and relative phases of all the NP parameters can be extracted from a
  fit to this angular distribution. One can measure CP-violating
  angular asymmetries. If one integrates over some of the five
  kinematic parameters parametrizing the angular distribution, one
  obtains (i) familiar observables such as the $q^2$ distribution and
  the $D^*$ polarization, and (ii) new observables associated with the
  $\pi^-$ emitted in the $\tau$ decay: the forward-backward asymmetry
  of the $\pi^-$ and the CP-violating triple-product asymmetry.}

\keywords{}

\arxivnumber{}

\preprint{
{\flushright
UdeM-GPP-TH-20-277 \\
UMISS-HEP-2020-01 \\
}}

\maketitle

\section{Introduction}

At the present time, there are discrepancies with the predictions of
the standard model (SM) in the measurements of some observables in a
number of $B$ decays.  These include $R_{D^{(*)}} \equiv \cB(\bar{B}
\to D^{(*)} \tau^{-} {\bar\nu}_\tau)/\cB(\bar{B} \to D^{(*)} \ell^{-}
    {\bar\nu}_\ell)$ ($\ell = e,\mu$) \cite{Lees:2012xj, Lees:2013uzd,
      Aaij:2015yra, Huschle:2015rga, Sato:2016svk, Hirose:2016wfn,
      Aaij:2017uff, Hirose:2017dxl, Aaij:2017deq, Abdesselam:2019dgh}
    and $R_{J/\psi} \equiv \cB(B_c^+ \to J/\psi\tau^+\nu_\tau) /
    \cB(B_c^+ \to J/\psi\mu^+\nu_\mu)$ \cite{Aaij:2017tyk}. The
    experimental results are shown in Table \ref{tab:obs_meas}. The
    values of the SM predictions for $R_D$ and $R_{D^*}$, as well as
    their experimental measurements, are the average values used by
    the Heavy Flavor Averaging Group (HFLAV) \cite{HFLAV}. They find
    that the deviation from the SM in $R_D$ and $R_{D^*}$ (combined)
    is $3.1\sigma$.\footnote{However, we note that this is not
      completely settled: for example, a more recent analysis finds
      $(R_{D^*}^{\tau/\ell})_{\rm SM} = 0.250 \pm 0.003$
      \cite{Bordone:2019vic}. With this value, not included in the
      HFLAV average, the deviation from the SM prediction is larger
      than $3.1\sigma$.} For $R_{J/\psi}$, the discrepancy with the SM
    is 1.7$\sigma$ \cite{Watanabe:2017mip}. These measurements suggest
    the presence of new physics (NP) in $\bctaunu$ decays.

\begin{table*}[h]
\begin{center}
\begin{tabular}{|c|c|c|}
 \hline\hline
Observable & SM Prediction & Measurement \\
\hline
$R_{D^*}^{\tau/\ell}$ & $0.258 \pm 0.005$ \cite{HFLAV} & $0.295 \pm 0.011 \pm 0.008$ \cite{HFLAV} \\
$R_{D}^{\tau/\ell}$ & $0.299 \pm 0.003$  \cite{HFLAV} & $0.340 \pm 0.027 \pm 0.013$ \cite{HFLAV} \\
$R_{J/\psi}^{\tau/\mu}$ & $0.283 \pm 0.048$ \cite{Watanabe:2017mip} & $0.71 \pm 0.17 \pm 0.18$ \cite{Aaij:2017tyk} \\
$R_{D^*}^{\mu/e}$ & $\sim 1.0$ & $1.04 \pm 0.05 \pm 0.01$ \cite{Abdesselam:2017kjf} \\
 \hline\hline
\end{tabular}
\end{center}
\caption{Measured values of observables that suggest NP in $\bctaunu$.}
\label{tab:obs_meas}
\end{table*}

A great many papers have examined the question of what type of NP is
required to explain the above anomalies. These include both
model-independent \cite{Watanabe:2017mip, Bhattacharya:2014wla,
Fajfer:2012jt, Datta:2012qk,Tanaka:2012nw, Biancofiore:2013ki, Freytsis:2015qca,
  Bardhan:2016uhr, Bhattacharya:2016zcw, Dutta:2017wpq,
   Alok:2017qsi, Chang:2018sud, Huang:2018nnq, Hu:2018veh, Alok:2019uqc, Kamali:2018fhr, Shivashankara:2015cta,
  Datta:2017aue}
  and
model-dependent analyses \cite{Crivellin:2012ye, Celis:2012dk,
  He:2012zp, Ko:2012sv, Dorsner:2013tla, Sakaki:2013bfa,
  Greljo:2015mma, Crivellin:2015hha, Dumont:2016xpj, Boucenna:2016wpr,
  Boucenna:2016qad, Bhattacharya:2016mcc, Alonso:2016oyd,
  Celis:2016azn, Wei:2017ago, Altmannshofer:2017poe,
  Matsuzaki:2017bpp, Buttazzo:2017ixm, Iguro:2017ysu, He:2017bft,
  Biswas:2018jun, Yang:2018pyq, Chen:2018hqy, Asadi:2018wea,
  Greljo:2018ogz, Abdullah:2018ets, Martinez:2018ynq, Fraser:2018aqj,
  Wang:2018upw, Kumar:2018kmr, Robinson:2018gza, Hu:2018lmk,
  Babu:2018vrl, Marzo:2019ldg, Aydemir:2019ynb, Yan:2019hpm,
  Gomez:2019xfw, Alok:2017sui,Alok:2017jgr, Datta:2019bzu,Datta:2019tuj, Beaudry:2017gtw,  Dev:2020qet, Crivellin:2019dwb,  Boubaa:2020ksf}.  Clearly there are many
possibilities for the NP. In order to distinguish the various NP
explanations, a variety of observables have been considered. These
include the $q^2$ distribution, the $D^*$ polarization, the $\tau$
polarization, etc.\ \cite{Datta:2012qk, Sakaki:2012ft, Sakaki:2014sea,
  Bhattacharya:2015ida, Alonso:2016gym, Alok:2016qyh, Ligeti:2016npd,
  Ivanov:2017mrj, Colangelo:2018cnj, Alok:2018uft, Azatov:2018knx,
  Asadi:2018sym, Iguro:2018vqb, Blanke:2018yud, Murgui:2019czp,
  Asadi:2019xrc, Blanke:2019qrx, Hill:2019zja, Ivanov:2020iad,
  Becirevic:2019tpx, Alguero:2020ukk, Mandal:2020htr}.

The above observables are all CP-conserving. But one can also consider
CP-violating observales in ${\bar B} \to D^{*} \tau^- {\bar\nu}_\tau$
\cite{Duraisamy:2013pia,Hagiwara:2014tsa, Duraisamy:2014sna, Aloni:2018ipm}.
All CP-violating effects require the interference of
two amplitudes with different weak (CP-odd) phases. Since the SM has
only one amplitude, the observation of CP violation in this decay
would be a smoking-gun signal of NP.

In Ref.~\cite{BDmunuCPV}, we began to explore the prospects for
measuring CP-violating effects in ${\bar B} \to D^{*} \tau^-
{\bar\nu}_\tau$.  There, we noted that, since ${\bar B} \to D^*$ is
the only hadronic transition in this decay, all amplitudes will have
the same strong (CP-even) phase. As a result, the direct CP asymmetry
is expected to be very small. The main CP-violating effects appear as
CP-violating asymmetries in the angular distribution. These are
kinematical observables, and require that the two interfering
amplitudes have different Lorentz structures.  This fact allows us to
distinguish different NP explanations. We demonstrated this by
constructing the angular distribution for the decay ${\bar B} \to
D^{*} \mu^- {\bar\nu}_\mu$, and showing that one could extract the
different NP contributions from an analysis of the CP-violating
angular asymmetries.

The reason we did not apply this to ${\bar B} \to D^{*} \tau^-
{\bar\nu}_\tau$ is that the construction of the angular distribution
requires the knowledge of the three-momentum ${\vec p}_\tau$. But
since the $\tau$ decays to final-state particles that include
$\nu_\tau$, which is undetected, ${\vec p}_\tau$ cannot be determined
with any precision.  As a result, the full angular distribution in
${\bar B} \to D^* (\to D \pi) \, \tau^{-} {\bar\nu}_\tau$ cannot be
measured\footnote{In fact, methods do exist that use all available
  experimental information to reconstruct the angular distribution.
  For example, Ref.~\cite{Hill:2019zja} uses the topology of decay
  vertices to perform a kinematic reconstruction. Still, in all of
  these methods, the angular distribution is obtained with limited
  precision (due to the uncertainty in the measurement of ${\vec
    p}_\tau$) and/or ambiguities.}.

In this paper, we construct a {\it measurable} angular distribution in
${\bar B} \to D^* (\to D \pi) \, \tau^{-} {\bar\nu}_\tau$. This is
obtained by considering the additional decay\footnote{We note in
  passing that the decay $\tau^- \to \pi^- \nu_\tau$ has been used in
  the context of a proposed method for measuring the $\tau$
  polarization in ${\bar B} \to D \tau^{-} (\to \pi^- \nu_\tau)
  {\bar\nu}_\tau$ \cite{Nierste:2008qe, Alonso:2017ktd}.} $\tau^- \to
\pi^- \nu_\tau$. Now there are three final-state particles whose
three-momenta {\it can} be measured: the $D$ and $\pi$ (from $D^*$
decay), and the $\pi^-$ (from $\tau$ decay).  The new angular
distribution is given in terms of five kinematic parameters: $q^2$,
$\theta^*$ (describing $D^* \to D\pi$), and three quantities
describing the $\pi^-$, $E_\pi$, $\theta_\pi$ and $\chi_\pi$. It
includes CP-violating angular asymmetries, which can be measured and
used to extract information about the NP.

But the angular distribution yields even more information. All the NP
parameters can be extracted from a fit to the full distribution. Thus,
even if the NP is CP-conserving, so that no CP-violating angular
asymmetries appear, its presence can still be detected. It is also
possible to integrate over one or more of the five parameters. When
one does this, all the familiar observables that have been proposed to
distinguish NP models, such as the $q^2$ distribution and the $D^*$
polarization, are reproduced. But there are also new observables that
depend on the kinematic angles associated with the $\pi^-$ emitted in
the $\tau$ decay, $\theta_\pi$ and $\chi_\pi$. These include the
forward-backward asymmetry of the $\pi^-$, and the CP-violating
triple-product asymmetry.

It should be noted that, in order to use this method, the momentum of
the decaying $B$ must be known. Thus, the technique described here is
more suited to the experiments at $e^+ e^-$ machines such as Belle II.

We begin in Sec.~2 with the derivation of the angular distribution of
$B \to D^{*}(\to D \pi^\prime) \tau(\to \pi \nu_\tau)
\bar{\nu}_{\tau}$. Here, some information is given in the Appendices.
In Sec.~3, we discuss the NP signals, both CP-conserving and
CP-violating, in the angular distribution. Observables obtained by
integrating this rate over one or more of the kinematical variables
are described in Sec.~4.  We conclude in Sec.~5.

\section{Angular Distribution}

We begin by describing our method of calculating the angular
distribution of $B \to D^{*}(\to D \pi^\prime) \tau(\to \pi \nu_\tau)
\bar{\nu}_{\tau}$. (Note that this section is somewhat technical. The
reader wishing to simply see the results may skip to the next
section.)

\subsection{Structure of the new angular distribution}

Consider first the angular distribution of the decay ${\bar B} \to D^*
(\to D \pi) \, \ell^{-} {\bar\nu}_\ell$. This is obtained as follows.
Assuming only left-handed (LH) neutrinos, the decay is parametrized as
${\bar B}\to D^*N^{*-}(\to\ell^-{\bar\nu}_\ell)$, where $N = S-P$,
$V-A$, $T$ represent LH scalar, vector and tensor interactions,
respectively. For $\ell = \mu, e$, there is no NP, so that $N = W$ and
the coupling is $V - A$. But for $\ell = \tau$, all couplings are
allowed. The full amplitude is then squared, and can be expressed as a
function of the final-state momenta. These momenta are defined in
terms of the three helicity angles of Fig.~1, $\theta_\ell$,
$\theta^*$ and $\chi$. In this way, one produces a set of angular
functions whose coefficients are different combinations of the
helicity amplitudes. This is the angular distribution
\cite{BDmunuCPV}.

\begin{figure}
\begin{center}
\includegraphics[width=0.6\textwidth]{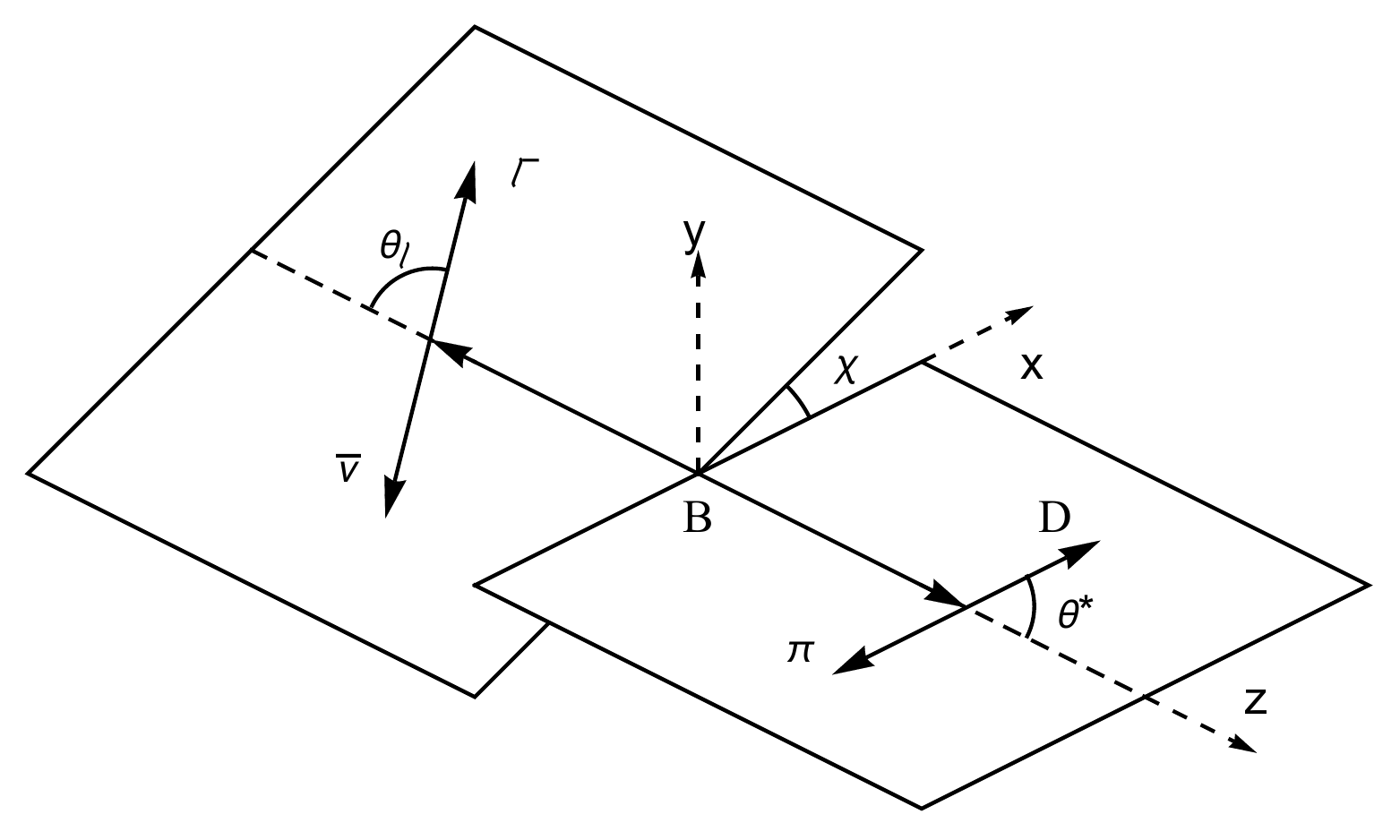}
\caption{Definition of the angles in the $\bar{B} \to D^* (\to D \pi) \, \ell^- \bar{\nu}_{\ell}$ distribution.}
\end{center}
\label{angles}
\end{figure}

We now consider the case where the final-state lepton is $\ell =
\tau$.  The $\tau$ is not directly detected in experiments; instead,
it is detected through its decay products. We choose to study the
simplest possible hadronic decay of the $\tau$, $\tau \to
\pi\nu_\tau$. While NP in the $\tau$ decay is a possibility,
in this analysis we restrict ourselves to NP only in the $B$ decay. As
we will show, even using this simple two-body decay of the $\tau$, one
can extract a great deal of information about this NP.

Once we let the $\tau$ decay, the process $B \to D^{*}(\to D
\pi^\prime) \tau(\to \pi \nu_\tau) \bar{\nu}_{\tau}$ has five
particles in the final state. This decay can be broken down into four
successive quasi-two-body decays of the $B$ meson and three
intermediate states. The five-body phase space for the decay of a
massive spinless particle, such as the $B$ meson, depends on 8
independent parameters: five helicity angles and the invariant squares
of the masses of the three intermediate particles. Since two of these
intermediates -- the $D^*$ and the $\tau$ -- can go on shell, two of
the three invariant mass parameters are given by $m_{D^*}$ and
$m_\tau$. Thus, this decay depends on six independent parameters: five
helicity angles and $q^2$, the invariant mass-squared of the
$\tau\bar{\nu}_\tau$ pair. In the following, given that it could be NP
that couples to $\tau\bar{\nu}_\tau$, we will refer to the
center-of-momentum frame of the $\tau\bar{\nu}_\tau$ pair as the $N^*$
rest frame.

Now, the helicity angles are typically defined in the rest frames of
the corresponding intermediate states. Following this procedure, we
define (i) $\theta^*$ as the polar angle of the $D$-meson
three-momentum in the rest frame of its parent $D^*$ meson, (ii)
$\theta_\tau$ and $\chi_\tau$ as the polar and azimuthal angles,
respectively, of the $\tau$ three-momentum in the $N^*$ rest frame,
and finally (iii) $\theta$ and $\chi$ as the polar and azimuthal
angles, respectively, of the $\pi$ three momentum in its parent $\tau$
rest frame.

However, this leads to a problem. Although one can in principle
theoretically define all five helicity angles, most of them are of no
practical use. To be specific, since the $\tau$ lepton is not directly
observed in experiments, the angles either associated with its three
momentum or defined in its rest frame are not measurable. Thus, four
of the five helicity angles ($\theta_{(\tau)}, \chi_{(\tau)}$) are of
no use to us. This problem can be remedied (at least partially)
through a convenient change of variables.

\subsection{New parameters}

Since we do not have experimental access to the $\tau$ rest frame, in
our analysis we choose to express the $\tau\to\pi\nu_\tau$ phase space
in the $N^*$ rest frame (this frame can be easily determined from
information about the hadronic side of the $B$ decay). Since the pion
three-momentum can be precisely measured in this frame, we consider
three new variables. $E_\pi$, $\theta_\pi$ and $\chi_\pi$ represent
the pion energy, polar and azimuthal angles, respectively, defined in
this frame. (The new helicity angles are shown in Fig.~2.)  These
three variables replace three of the unmeasurable helicity angles. The
fourth unmeasurable angle is an azimuthal angle and is easily
integrated over. We describe below the mathematical method for this
transformation.

\begin{figure}
\begin{center}
\includegraphics[width=0.6\textwidth]{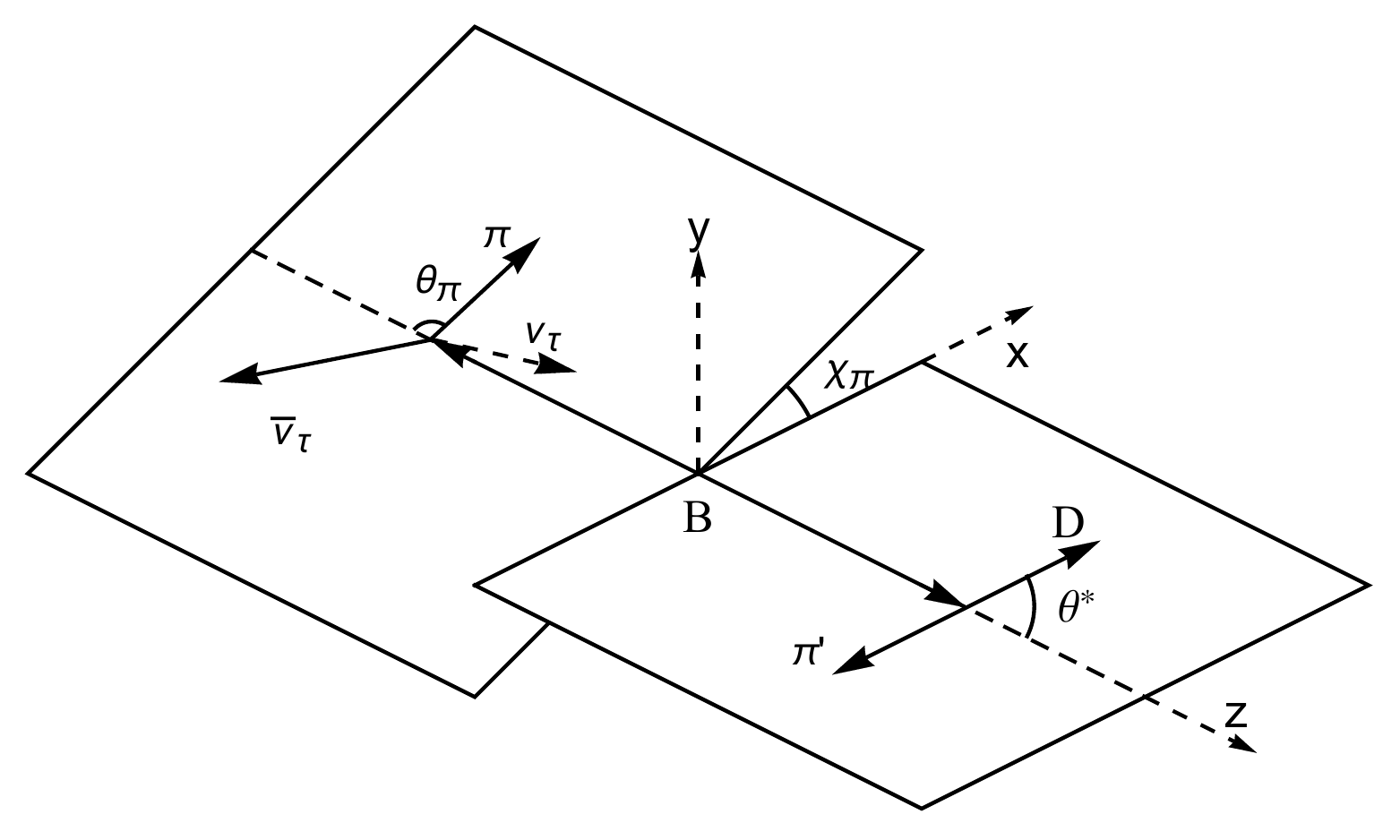}
\caption{Definition of the angles in the $\bar{B} \to D^* (\to D \pi)
  \, \tau^- (\to \pi^- \nu_\tau) \bar{\nu}_{\tau}$ distribution.}
\end{center}
\label{fig:angles}
\end{figure}

Let us consider the product, $d^4I$, of the quasi-two-body phase
spaces for $N^* \to \tau \bar{\nu}_\tau$ ($\phi_{N^*}$) and $\tau \to
\pi \nu_\tau$ ($\phi_\tau$). (The $d^4$ serves as a reminder that this
phase-space factor ultimately depends on four independent kinematic
variables.) Each phase-space factor is evaluated in the corresponding
parent rest frame, and is expressed in terms of the four unmeasurable
helicity angles. However, since each individual phase-space factor is
Lorentz invariant, we can write this entire product in the measurable
$N^*$ rest frame:
\bea
d^4I &=& \int d\phi_{N^*}(p_\tau,p_{\nutab}) \int d\phi_\tau(p_\pi,p_{\nuta}) ~,~~ \nl
  &=& \frac{1}{(4\pi)^{4}}\int \frac{d^3p_\tau d^3p_{\nutab}}{E_\tau E_{\nutab}}
   \, \de^4(q - p_\tau - p_{\nutab}) \int \frac{d^3p_\pi d^3p_{\nuta}}{E_\pi E_{\nuta}}
  \de^4(p_\tau - p_\pi - p_{\nuta}) ~,~~ \nl
  &=& \frac{1}{(4\pi)^{4}}\int \frac{d^3p_\tau d^3p_{\nutab}}{E_\tau E_{\nutab}}
   \, \de\(\sqrt{q^2} - E_\tau - E_{\nutab}\)\de^3\(\vp_\tau + \vp_{\nutab}\) ~~~ \nl
  && \hspace{3truecm} \int \frac{d^3p_\pi d^3p_{\nuta}}{E_\pi E_{\nuta}} \, \de\(E_\tau
  - E_\pi - E_{\nuta}\)\de^3\(\vp_\tau - \vp_\pi - \vp_{\nuta}\) ~,~~
\eea
where, in the final line, $E_x$ and $\vp_x$ respectively represent the
energy and three-momentum of the particle $x$ in the $N^*$ rest
frame. Performing the integrals over the $\nuta$ and $\nutab$
three-momenta, and neglecting neutrino masses, we find
\bea
d^4I &=& \frac{1}{(4\pi)^{4}}\int\frac{d^3p_\tau}{E_\tau|\vp_\tau|} \, \de(\sqrt{q^2} -
E_\tau - |\vp_\tau|)\int\frac{d^3p_\pi}{E_\pi|\vp_\tau - \vp_\pi|} \, \de(E_\tau - E_\pi
- |\vp_\tau - \vp_\pi|) ~.~~
\eea

Without loss of generality, we now choose to write the $\tau$ and
$\pi$ three-momentum integral measures such that the associated polar
angle can be determined, at least theoretically. In the case of
$d^3p_\pi$, clearly the polar and azimuthal angles of the pion
three-momentum relative to the $N^*$ direction, $\theta_\pi$ and
$\chi_\pi$ respectively, are measurable. Here, $\theta_\pi$ is
defined using three-momenta evaluated in the $N^*$ rest frame,
\beq
\cos \theta_\pi = - \frac{\vp_{D^*} \cdot \vp_\pi}{|\vp_{D^*}||\vp_\pi|} ~,
\eeq
while $\chi_\pi$ is defined using three-momenta evaluated in the
$B$ rest frame,
\beq
\sin\chi_\pi = \frac{\[(\vp_{\pi'} \times \vp_D) \times (\vp_{D^*} \times
\vp_\pi)\]\cdot\vp_{D^*}}{|\vp_{\pi'}\times\vp_D||\vp_{D^*}\times\vp_\pi|
|\vp_{D^*}|} ~.
\eeq
Since $\vp_{D^*} = \vp_D + \vp_{\pi'}$, one can easily verify that
$\sin \chi_\pi$ is proportional to the scalar triple product
$(\vp_{\pi'} \times \vp_D) \cdot \vp_\pi$.

In the case of $d^3p_\tau$, the polar angle of the $\tau$ direction
relative to the pion direction, $\theta_{\tau\pi}$, can be
theoretically determined. The fourth angle -- the corresponding
azimuthal angle $\chi_{\tau\pi}$ -- cannot be determined. However, at
a later stage we will eliminate this angle by integrating over
it. After appropriately transforming the delta functions, and writing
the phase space in terms of the above new variables ($\theta_\pi$,
$\chi_\pi$, $\theta_{\tau\pi}$ and $\chi_{\tau\pi}$), we find
\bea
d^4I &=& \frac{1}{(4\pi)^{4}}\int\frac{d|\vp_\tau|}{\sqrt{q^2}} \, d\cos\theta_{\tau\pi}
 \, d\chi_{\tau\pi} \, dE_\pi  \, d\cos\theta_\pi  \, d\chi_\pi ~~~ \nl
&& \hspace{3truecm} \de\(|\vp_\tau| - \frac{q^2 - m^2_\tau}{2\sqrt{q^2}}\)~\de\(\cos
\theta_{\tau\pi} - \frac{2E_\tau E_\pi - m^2_\tau - m^2_\pi}{2|\vp_\tau||\vp_\pi|}\) ~.~~
\eea

Expressed in the above form, it is clear that the remaining two delta
functions can be used to remove the two variables $|\vp_\tau|$ and
$\cos\theta_{\tau\pi}$. We are thus left with a phase-space factor
that depends only on four variables ($\chi_{\tau\pi}$, $E_\pi$,
$\theta_\pi$ and $\chi_\pi$), as expressed below:
\bea
d^4I &=& \frac{1}{(4\pi)^{4}} \, \frac{1}{\sqrt{q^2}} \, d\chi_{\tau\pi} \, dE_\pi  \, d\cos\theta_\pi \, d\chi_\pi ~,~~
\eea
where the following replacements in the squared invariant amplitude of
the decay ($|\cM|^2$) are understood:
\beq
E_\tau \to \frac{q^2 + m^2_\tau}{2\sqrt{q^2}} ~~,~~~~
|\vp_\tau| \to \frac{q^2 - m^2_\tau}{2\sqrt{q^2}} ~~,~~~~
\cos\theta_{\tau\pi} \to \frac{2E_\tau E_\pi - m^2_\tau - m^2_\pi}{2|\vp_\tau||\vp_\pi|} ~.~~
\eeq
Using the above choice of kinematic parameters we may now express the
differential decay rate for the full process as follows:
\bea
\frac{d^5\Ga}{dq^2 \, d\cos\theta^* \, dE_\pi \, d\cos\theta_\pi \, d\chi_\pi} &=& \frac{|\vp_{D^*}| \, |\vp_{D}|}
{2^{15} \, \pi^7 \, m^2_B \, m_{D^*} \, \sqrt{q^2}}\int d\chi_{\tau\pi}  \, \frac{dp^2_{D^*}}{2\pi} \, \frac{dp^2_\tau}
{2\pi}|\cM|^2 ~. \label{eq:diffrate}
\eea
Here $|\vp_{D^*}| = \sqrt{\la(m^2_B;q^2,m^2_{D^*})}/(2 m_B)$ and $|\vp_D| = \sqrt{\la(m^2_{D^*};m^2_D,
m^2_\pi)}/(2 m_{D^*})$, where
\bea
\la(a;b,c) &=& a^2 + b^2 + c^2 - 2 a b - 2 b c - 2 c a ~.~~
\eea
The right-hand side of Eq. (\ref{eq:diffrate}) contains integrals over
three independent variables (out of the eight variables discussed in
the previous subsection). We will see in the following subsection that
these integrals can be performed quite simply once we express $|\cM|^2$
as an explicit function of these variables.

\subsection{Calculating $|\cM|^2$}
\label{sec:msq}

The next step is to calculate $|\cM|^2$, appropriately summed over
spins and polarizations. In Ref.~\cite{BDmunuCPV}, we derived the
angular distribution for $B\to D^*\mu \bar{\nu}_\mu$. In the presence
of NP, the relevant two-body processes are ${\bar B}\to
D^*N^{*-}(\to\mu^-{\bar\nu}_\mu)$, where $N = S-P, V-A, T$ represent
left-handed scalar, vector and tensor interactions,
respectively. These are labeled $SP$, $VA$ and $T$.  (The $VA$
contribution includes that of the SM.) For each of the leptonic $SP$,
$VA$ and $T$ Lorentz structures, the hadronic piece (the $b \to c$
transition) also has a NP contribution. The effective Hamiltonian is
\bea
{\cal H}_{eff} &=& \frac{G_F V_{cb}}{\sqrt{2}} \Bigl\{
\left[ (1 + g_L) \, {\bar c} \gamma_\al (1 - \gamma_5) b + g_R \, {\bar c} \gamma_\al (1 + \gamma_5) b \right]
{\bar \mu} \gamma^\al (1 - \gamma_5) \nu_\mu \nn\\
&& \hskip-1truecm
+~\left[ g_S \, {\bar c} b + g_P \, {\bar c} \gamma_5 b \right] {\bar \mu} (1 - \gamma_5) \nu_\mu
+ g_T \, {\bar c} \sigma^{\al\be} (1 - \gamma_5) b
{\bar \mu} \sigma_{\al\be} (1 - \gamma_5) \nu_\mu\Bigr\} + h.c.
\label{4fermi_NP}
\eea
The decay amplitude is then written as the product of a hadronic piece
$\cH_{D^*}$, a leptonic piece $\cL^{N^*}$, and a helicity amplitude
piece $\cM^{N^*}$, appropriately summed over helicities labeled by
$m,n$, and $p$.

This all applies to the decay $B\to D^*\tau \bar{\nu}_\tau$, except
that now one must also include the decay $\tau\to\pi\nuta$. In
addition to numerical factors and factors of $f_\pi|V_{ud}|$ coming
from the $\tau\to\pi\nuta$ transition, the leptonic piece changes.
Representing the new leptonic pieces by $\tL^{N^*}$, the spin-summed
squared invariant amplitude for the full 5-body decay can now be
expressed as
\bea
|\cM|^2 &=& \frac{96\pi \, G_F^2 \, |V_{cb}|^2 \, m_{D^*}}{|\vp_D|^3 \,
(m^2_\tau - m^2_\pi)^2} \, \frac{m_{D^*}\,\Ga_{D^*}\,\cB(D^*\to D\pi')}
{(p^2_{D^*} - m^2_{D^*})^2 + m^2_{D^*}\Ga^2_{D^*}} \, \frac{m_\tau \,\Ga_\tau
\, \cB(\tau\to\pi\nuta)}{(p^2_\tau - m^2_\tau)^2 + m^2_\tau\Ga^2_\tau} ~~~ \nl
&&\hspace{5truemm}\lp\sum\limits_{m=\pm,0}\cH_{D^*}(m)\(\cM_{(m)}^{SP}\,\tL
^{SP}+ \sum\limits_{n=t,\pm,0}g_{nn}\cM_{(m;n)}^{VA}\tL^{VA}(n)\rd\rd ~ \nl
&&\hspace{5truecm} \ld\ld + \sum\limits_{n,p=t,\pm,0}g_{nn}g_{pp}\cM_{(m;n,
p)}^T\tL^T(n,p)\)\rp^2 ~.
\eea
In the above, the new leptonic pieces are of the form
\bea
\tL^{SP} &=& m_\tau\bar{u}(\nuta) \sp_\pi (1 - \ga_5) v(\nutab) ~,  \nl
\tL^{VA}(n) &=& \ep_{VA}^{\beta}(n)\[\bar{u}(\nuta) \sp_\pi \sp_\tau
\ga_\be (1 - \ga_5) v(\nutab) \] ~, \nn\\
\tL^{T}(n,p) &=& -i\,m_\tau\,\ep_{T}^{\be}(n)\ep_{T}^{\de} (p) \, \[\bar{u}(\nuta)
\sp_\pi \si_{\be\de} (1 - \ga_5) v(\nutab)\] ~,
\eea
and we have used the SM expressions for the branching fractions
$\cB(D^*\to D\pi')$, and $\cB(\tau\to\pi\nuta)$:
\beq
\cB(\tau\to\pi\nuta) ~=~ \frac{G^2_F \, |V_{ud}|^2 \, f^2_\pi}{16\pi\, m_\tau\,
\Ga_\tau}(m_\tau^2 - m_\pi^2)^2 ~~,~~~~
\cB(D^*\to D\pi') ~=~ \frac{|\vp_{D}|^3}{6\pi\,m^2_{D^*}\,\Ga_{D^*}}~.
\eeq
The hadronic pieces, $\cH_{D^*}$, and the helicity amplitude pieces $\cM^{N^*}$
are the same as those obtained in our earlier work, Ref.\ \cite{BDmunuCPV}.
For completeness, we have provided this information in Appendix \ref{app:helicity}.

We now see that the dependence of $|\cM|^2$ on the variables
$p_{D^*}^2$ and $p^2_\tau$ appears only through the propagators of
the corresponding intermediate particles. Since both of these
particles -- the $D^*$ and the $\tau$ -- go on shell, we can apply the
narrow-width approximation to replace these propagators with delta
functions, making the corresponding integrals simple. Under the
narrow-width approximation, one can show that
\bea
\int\frac{dp^2}{2\pi}\frac{m_X\,\Ga_X\,\cB}{(p^2 - m_X^2)^2 + m_X^2\Ga_X^2} &\to& \frac{\cB}{2} ~.
\eea
Furthermore, the dependence of $|\cM|^2$ on the unmeasurable azimuthal
angle $\chi_{\tau\pi}$ is a result of fermionic traces over products
of the leptonic pieces ($\tL^{N^*}$). This dependence turns out to be
combinations of simple trigonometric functions, such as
$\sin\chi_{\tau\pi}$ and $\cos\chi_{\tau\pi}$. It is therefore
straightforward to integrate over $\chi_{\tau\pi}$.

After integrating over the three variables $p^2_{D^*}, p^2_\tau$ and
$\chi_{\tau\pi}$, the full five-body differential decay rate is given
by
\bea
\frac{d^5\Ga}{dq^2\,dE_\pi\,d\cos\theta^*\,d\cos\theta_\pi\,d\chi_\pi} &=& \frac{3|V_{cb}|^2\,
G_F^2\,|\vec{p}_{D^*}|\,(q^2)^{3/2}\,m_\tau^2}{2^{11}\pi^4 m_B^2 (m_\tau^2-m_\pi^2)^2} \,
\cB(D^*\to D\pi')\cB(\tau\to\pi\nuta) \nl
&&\hspace{-1truecm}\times\sum_{i,j} \(\cN_i^S \, |\cA_i|^2 + \cN_{i,j}^R \, {\rm Re}[\cA_i\cA_j^*] +
\cN_{i,j}^I \, {\rm Im}[\cA_i \cA_j^*]\) ~,
\label{eq:angdist}
\eea
where $i,j = t, 0, \bot, \parallel, SP, (0,T), (\bot,T),
(\parallel,T)$.  Here the $\cA_i$ represent the helicity amplitudes
that contain the crucial physics information that can be extracted
from this analysis, while the $\cN^{(S,R,I)}_{i(,j)}$ are functions of
the five independent kinematic variables of interest to us
($q^2,\theta^*,E_\pi,\theta_\pi$, and $\chi_\pi$).
We present the information relevant for the $\cN_i^S \,|\cA_i|^2$
pieces of Eq.~(\ref{eq:angdist}) in Table \ref{tab:NS} of Appendix
B. The first column contains the various $|\cA_i|^2$ helicities, while
the second column contains the associated $\cN_i^S$ terms. In these
terms, we have separated out the parts that depend on $q^2$ and
$E_\pi$, and put them into the $S_i$ factors. The expressions for the
$S_i$ are also given in Appendix B. The information relevant for the
$\cN_{i,j}^R \, {\rm Re}[\cA_i\cA_j^*]$ and $\cN_{i,j}^I \, {\rm
  Im}[\cA_i \cA_j^*]$ pieces is given in Tables \ref{tab:NR} and
\ref{tab:NI} of Appendix B, respectively. The expressions for the
$R_i$ and $I_i$ are also given in Appendix B.

It is standard to express the differential decay rate as an angular
distribution, written as a sum over a product of angular functions and
functions of non-angular variables including the helicity amplitudes.
In order to write Eq.\ (\ref{eq:angdist}) as an angular distribution,
it is necessary to separate the $\cN^{(S,R,I)}_{i(,j)}$ into angular
functions and functions of $q^2$ and $E_\pi$. Following this
separation, Eq.\ (\ref{eq:angdist}) can be rewritten as a sum over a
product of 12 angular functions and their respective coefficients:
\bea
\frac{d^5\Ga}{dq^2\,dE_\pi\,d\cos\theta^*\,d\cos\theta_\pi\,d\chi_\pi}
&=& \frac{3|V_{cb}|^2\,G_F^2\,|\vec{p}_{D^*}|\,(q^2)^{3/2}\,m_\tau^2}
{2^{11}\pi^4 m_B^2 (m_\tau^2-m_\pi^2)^2} \, \cB(D^*\to D\pi')\cB(\tau
\to\pi\nuta) \nl
&&\hspace{-3truecm}\times\[\sum^{9}_{i=1} f^R_i(q^2,E_\pi)\Omega^R_i(\theta^*,
\theta_\pi,\chi_\pi) + \sum^{3}_{i = 1}f^I_i(q^2,E_\pi)\Omega^I_i(\theta^*,
\theta_\pi,\chi_\pi)\] ~.
\label{eq:angdist2}
\eea
The first nine angular functions above, denoted by $\Omega^R_{1,\ldots 9}$,
arise from a rearrangement of the $\cN_i^S \, |\cA_i|^2 + \cN_{i,j}^R \,
{\rm Re}[\cA_i\cA_j^*]$ terms. These are presented in Table \ref{tab:NSR}.
The $\cN_{i,j}^I \,{\rm Im}[\cA_i\cA_j^*]$ terms involve an additional three
angular functions, denoted by $\Omega^I_{1,2,3}$; these contributions are
given in Table \ref{tab:NI2}. Together, these constitute the ${\bar B} \to
D^* (\to D \pi') \, \tau^{-} (\to \pi^- \nu_\tau) {\bar\nu}_\tau$ angular
distribution.

\begin{table}
\begin{center}
\begin{tabular}{|c|c|} \hline
Coefficient       & Angular Function\\
 $f^R_i(q^2,E_\pi)$ &  $\Omega^R_i(\theta^*,\theta_\pi,\chi_\pi)$ \\ \hline \hline
$S_t\lp\cA_t\rp^2 + S_{0,1}\lp\cA_0\rp^2 + S_{SP}\lp\cA_{SP}\rp^2 + S_{0T,1}\lp\cA_{0,T}\rp^2$ &\\
 $+ R_{0T01} \, \Re[\cA_{0,T} \cA_0^*] + R_{SPt} \, \Re[\cA_{SP}\cA_t^*] $
     &   $\cos^2\theta^*$ \\ \hline
$S_{\bot ,1}\lp\cA_{\bot }\rp^2 +  S_{\parallel,1} \lp\cA_{\parallel }\rp^2 + S_{\bot T,1}\lp\cA_{\bot ,T}\rp^2 + S_{\parallel T,1}\lp\cA_{\parallel ,T}\rp^2 $ & \\
   $+ R_{\parallel T \parallel,1}  \, \Re[\cA_{\parallel,T}\cA_{\parallel}^*] + R_{\bot T \bot,1}  \, \Re[\cA_{\bot,T}\cA_{\bot}^*] $
    &    $\sin^2\theta^*$  \\ \hline
$R_{SP0} \, \Re[\cA_{SP}\cA_{0}^*]+R_{t0} \, \Re[\cA_{t}\cA_{0}^*]$ & \\
$+R_{SP0T} \, \Re[\cA_{SP}\cA_{0,T}^*] +R_{0Tt} \, \Re[\cA_{0,T}\cA_{t}^*] $        &    $\cos^2\theta^* \, \cos\theta_\pi$ \\ \hline
$S_{0,2}\lp\cA_0\rp^2 + S_{0T,2}\lp\cA_{0,T}\rp^2 + R_{0T0,2} \, \Re[\cA_{0,T}\cA_{0}^*] $        &    $\cos^2\theta^* \, \cos2\theta_\pi$ \\ \hline
$ R_{\bot T \parallel} \, \Re[\cA_{\bot,T}\cA_{\parallel}^*] + R_{\bot T \parallel T} \, \Re[\cA_{\bot, T}\cA_{\parallel,T}^*]$ &\\
$+R_{\parallel \bot} \, \Re[\cA_{\parallel}\cA_{\bot}^*] + R_{\parallel T \bot} \, \Re[\cA_{\parallel,T}\cA_{\bot}^*]$        &    $\sin^2\theta^* \, \cos\theta_\pi$ \\ \hline
$ S_{\parallel,2} \lp\cA_{\parallel}\rp^2 + S_{\bot,2}\lp\cA_{\bot }\rp^2 + S_{\parallel T,2}\lp\cA_{\parallel ,T}\rp^2 + S_{\bot T,2}\lp\cA_{\bot ,T}\rp^2 $ & \\
$+ R_{\parallel T \parallel,2} \, \Re[\cA_{\parallel,T}\cA_{\parallel}^*] + R_{\bot T \bot,2} \, \Re[\cA_{\bot,T}\cA_{\bot}^*] $        &    $\sin^2\theta^* \, \cos2\theta_\pi$ \\ \hline
$ 2 (S_{\bot,2}\lp\cA_{\bot }\rp^2 - S_{\parallel,2} \lp\cA_{\parallel}\rp^2) + 2 (S_{\bot T,2}\lp\cA_{\bot ,T}\rp^2 -  S_{\parallel T,2}\lp\cA_{\parallel ,T}\rp^2)$ & \\
 $+ 2 (R_{\bot T \bot,2} \, \Re[\cA_{\bot,T}\cA_{\bot}^*] - R_{\parallel T \parallel,2} \, \Re[\cA_{\parallel,T}\cA_{\parallel}^*]) $        &    $\sin^2\theta^* \, \sin^2\theta_\pi \cos2\chi_\pi$ \\ \hline
$R_{\bot T 0} \, \Re[\cA_{\bot,T}\cA_{0}^*] + R_{SP \parallel} \, \Re[\cA_{SP}\cA_{\parallel}^*]
+ R_{t \parallel} \, \Re[\cA_{t}\cA_{\parallel}^*]$ &  \\
$+ R_{SP \parallel T} \, \Re[\cA_{SP}\cA_{\parallel,T}^*] $
  $+ R_{0 \bot} \, \Re[\cA_{0}\cA_{\bot}^*]+ R_{0T\bot} \, \Re[\cA_{0,T}\cA_{\bot}^*] $ & \\
$+ R_{0T\bot T} \, \Re[\cA_{0,T}\cA_{\bot,T}^*] + R_{\parallel T t} \, \Re[\cA_{\parallel,T}\cA_{t}^*]  $        &    $\sin2\theta^* \, \sin\theta_\pi \, \cos\chi_\pi$ \\ \hline
$R_{\parallel T 0} \, \Re[\cA_{\parallel,T}\cA_{0}^*] + R_{0\parallel} \, \Re[\cA_{0}\cA_{\parallel}^*]$ & \\
$+ R_{0T \parallel} \, \Re[\cA_{0,T}\cA_{\parallel}^*] + R_{0T \parallel T} \, \Re[\cA_{0,T}\cA_{\parallel,T}^*] $        &    $\sin2\theta^* \, \sin2\theta_\pi \, \cos\chi_\pi$ \\ \hline
\end{tabular}
\caption{Contributions of $\cN_i^S \, |\cA_i|^2$ and $\cN_{i,j}^R \,
  {\rm Re}[\cA_i\cA_j^*]$ [Eq.\ (\ref{eq:angdist})] to the angular
  distribution. These terms are CP-conserving.
\label{tab:NSR}}
\end{center}
\end{table}

\begin{table}[h]
\begin{center}
\begin{tabular}{|c|c|} \hline
Coefficient       & Angular Function\\
 $f^I_i(q^2,E_\pi)$ &  $\Omega^I_i(\theta^*,\theta_\pi,\chi_\pi)$ \\ \hline
\hline
$I_{t\bot} \, \Im[\cA_t \cA_{\bot}^*] + I_{\parallel T0} \, \Im[\cA_{\parallel ,T} \cA_0^*]+ I_{SP\bot} \, \Im[\cA_{SP}\cA_{\bot}^*]$ & \\
$+ I_{SP\bot T} \, \Im[\cA_{SP}\cA_{\bot,T}^*]+ I_{0T\parallel} \, \Im[\cA_{0,T} \cA_{\parallel }^*] + I_{\bot Tt} \, \Im[\cA_{\bot ,T} \cA_t^*]$    & $\sin2\theta^* \sin\theta_\pi \sin\chi_\pi$  \\ \hline
$I_{0\bot}  \, \Im[\cA_0 \cA_{\bot}^*]+ I_{0T\bot} \, \Im[\cA_{0,T} \cA_{\bot }^*]+ I_{\bot T0} \, \Im[\cA_{\bot ,T} \cA_0^*]$ & $\sin2\theta^* \sin2\theta_\pi \sin\chi_\pi$ \\ \hline
$I_{\parallel \bot } \, \Im[\cA_{\parallel}\cA_{\bot}^*]+ I_{\bot T\parallel}  \, \Im[\cA_{\bot ,T} \cA_{\parallel }^* ]+I_{\parallel T\bot}  \, \Im[\cA_{\parallel ,T} \cA_{\bot }^*]$ & $\sin^2\theta^* \sin^2\theta_\pi \sin 2\chi_\pi$ \\  \hline
\end{tabular}
\caption{Contributions of $\cN_{i,j}^I \, {\rm Im}[\cA_i \cA_j^*]$
  [Eq.\ (\ref{eq:angdist})] to the angular distribution. These terms
  are CP-violating.
\label{tab:NI2}}
\end{center}
\end{table}

\section{Angular Distribution: New-Physics Signals}

Tables \ref{tab:NSR} and \ref{tab:NI2} describe the angular
distribution of the decay ${\bar B} \to D^* (\to D \pi') \, \tau^-
(\to\pi^-\nuta)\nutab$. The question now is: how can we use it to
obtain information about NP? This is discussed in the present section.

In decays such as $B \to K^* (\to K \pi) \mu^+ \mu^-$, where the only
non-angular parameter is $q^2$, the data is separated into $q^2$ bins
before an angular analysis is performed. In the present case, the
situation is similar, except that there are two non-angular variables,
$q^2$ and $E_\pi$. Therefore, the data has to be separated into both
$q^2$ and $E_\pi$ bins. An angular fit to the data can then be performed,
permitting the extraction of the coefficients in Tables \ref{tab:NSR}
and \ref{tab:NI2}.

These coefficients involve the eight helicity amplitudes $\cA_0$,
$\cA_\parallel$, $\cA_\bot$, $\cA_t$, $\cA_{SP}$, $\cA_{0,T}$,
$\cA_{\parallel ,T}$ and $\cA_{\bot ,T}$. In Eq.~(\ref{4fermi_NP}),
there are five NP parameters: $g_L$, $g_R$, $g_S$, $g_P$ and $g_T$. Of
these, $g_S$ does not contribute to this decay. The eight helicity
amplitudes are generated, at least in part, by the remaining four NP
parameters. The dependence of the helicity amplitudes on the NP
parameters is shown in Table \ref{tab:couplings}. Note that the
Lorentz structure associated with $g_L$ is $(V-A)\times(V-A)$, as in
the SM. For this reason, it is the quantity $1 + g_L$ that appears in
the Table, where the $1$ is due to the SM. Thus, $\cA_0$,
$\cA_\parallel$, $\cA_\bot$ and $\cA_t$ are present in the SM -- they
are associated with $W$ exchange -- while $\cA_{SP}$, $\cA_{0,T}$,
$\cA_{\parallel ,T}$ and $\cA_{\bot ,T}$ are purely NP helicity
amplitudes.

\begin{table}[h]
\begin{center}
\begin{tabular}{|c|c|} \hline
Helicity Amplitude & Coupling \\
\hline
$\cA_0$, $\cA_\parallel$, $\cA_t$ & $1+g_L-g_R$ \\
$\cA_\bot$ & $1+g_L+g_R$ \\
$\cA_{SP}$ & $g_P$ \\
$\cA_{0,T}$, $\cA_{\parallel ,T}$, $\cA_{\bot ,T}$ & $g_T$ \\
\hline
\end{tabular}
\caption{Contributions of the NP couplings to the various helicity amplitudes.
\label{tab:couplings}}
\end{center}
\end{table}

The couplings $g_L$, $g_R$, $g_P$ and $g_T$ are complex quantities, i.e. each
coupling has an independent magnitude and a weak (CP-odd) phase. The amplitudes
$\cA_i$ are constructed by taking a product of a coupling with a corresponding
QCD matrix element, and a numerical factor that appears in the effective
Hamiltonian of Eq. (\ref{4fermi_NP}). In principle, each QCD matrix element has
a strong (CP-even) phase. This means that, in principle, amplitudes
may have both weak (CP-odd) and strong (CP-even) phases. However, as argued in
Refs.~\cite{Datta:2004re, Datta:2004jm, BDmunuCPV} (and summarized in the
introduction), we expect all amplitudes to have the same strong phase as that
of the SM.

The coefficients in Tables \ref{tab:NSR} and \ref{tab:NI2} involve
products of the helicity amplitudes: $\lp\cA_i\rp^2$, $\Re[\cA_i
  \cA_j^*]$, $\Im[\cA_i \cA_j^*]$. With the above assumption, these
products can all be written in terms of seven NP parameters: the four
magnitudes of $1 + g_L$, $g_R$, $g_P$ and $g_T$, and their three
relative weak phases. Thus, the measurement of the angular
distribution allows us to probe these NP parameters.

If all NP quantities have the same weak phase, then $\Im[\cA_i
  \cA_j^*] = 0$, so that all the entries of Table \ref{tab:NI2}
vanish. On the other hand, those of Table \ref{tab:NSR} do not. For
this reason, we refer to Table \ref{tab:NSR} as CP-conserving, and
Table \ref{tab:NI2} as CP-violating.

Note that, if the strong-phase differences are nonzero, this is not
completely accurate. With nonzero strong-phase differences, the
entries of Tables \ref{tab:NS} and \ref{tab:NR} can differ between
${\bar B} \to D^{*} \tau^- {\bar\nu}_\tau$ and its CP-conjugate
process. That is, there can be direct CP violation. However, if an
untagged data sample is used to measure the angular distribution,
i.e., both process and CP-conjugate process are combined, then Table
\ref{tab:NSR} is indeed CP-conserving. As for Table \ref{tab:NI2}, its
entries are CP-violating and can be nonzero even in the untagged data
sample (details are given below).

In the following subsections, we examine how to obtain NP information
from the measurement of Tables \ref{tab:NI2} and \ref{tab:NSR}. As we
will see, Table \ref{tab:NI2} provides smoking-gun signals of NP,
while more work is require to identify NP in Table \ref{tab:NSR}.

\subsection{CP-violating angular terms}

Above, we argued that the strong-phase differences between the various
amplitudes are expected to be very small. This implies that all direct
CP-violating effects are also expected to be tiny. Even so,
CP-violating effects can be present in the angular distribution. To be
specific, the coefficients of certain angular terms are related to
triple products (TPs) of the form ${\vec p}_1 \cdot ({\vec p}_2 \times
{\vec p}_3)$, where the ${\vec p}_i$ are the final-state momenta. As
we will see below, TP asymmetries do not require a strong-phase
difference between the interfering amplitudes. Indeed, they are
maximal when this strong-phase difference vanishes. In the decay $B
\to D^{*}(\to D \pi^\prime) \tau(\to \pi \nu_\tau) \bar{\nu}_{\tau}$,
the 3-momenta of the final-state particles $D$, $\pi$ and $\pi^\prime$
can be measured.  From these, a TP can be constructed; all the entries
of Table \ref{tab:NI2} involve this TP.

Now, all entries are proportional to ${\rm Im}[A_i A_j^*]$, where
$A_i$ and $A_j$ are the two interfering helicity amplitues. Writing
\beq
A_i = |A_i| e^{i\phi_i} e^{i\delta_i} ~~,~~~~ A_j = |A_j| e^{i\phi_j} e^{i\delta_j} ~,
\eeq
where $\phi_{i,j}$ ($\delta_{i,j}$) are the weak (strong) phases, we see that
\beq
{\rm Im}[A_i A_j^*] = |A_i| |A_j| \sin(\phi_i - \phi_j + \delta_i - \delta_j) ~.
\eeq
If, as we have assumed, the strong-phase difference is negligible, the
TP is proportional to $\sin(\phi_i - \phi_j)$. This is a CP-violating
quantity. On the other hand, if the strong-phase difference is not
negligible, the TP can be nonzero even if the weak-phase difference
vanishes. That is, this is not CP-violating (it is known as a ``fake
TP''). To obtain a true CP-violating term, this must be compared to
the TP in the CP-conjugate process. In the CP-conjugate process, the
weak phases change sign, but the strong phases do not. But there is an
additional change.  Each angular function in Table \ref{tab:NI2} is
proportional to $\sin\chi_\pi$, so that these functions are parity
odd. This means that, in going from process to CP-conjugate process,
there is an additional minus sign \cite{Datta:2003mj,Gronau:2011cf},
so that the CP-conjugate TP is proportional to
\beq
-{\rm Im}[{\bar A}_i {\bar A}_j^*] = |A_i| |A_j| \sin(\phi_i - \phi_j - \delta_i + \delta_j) ~.
\eeq
The true, CP-violating effect is then found by {\it adding} the TPs in
process and CP-conjugate process \cite{Datta:2003mj}, so that it remains
even in an untagged data sample\footnote{Whether to add or subtract individual
angular terms for the construction of a true CP-violating effect depends on the
sign convention used to define the azimuthal angle. Theory sign conventions for
the decay $B\to K^*\mu^-\mu^+$, which our discussion follows for the $B\to D^*
\tau^-\nutab$, can be found in Ref. \cite{Altmannshofer:2008dz}. Ref.
\cite{Gratrex:2015hna} presents detailed comparisons between sign conventions
used in  $B\to K^*\mu^+\mu^-$ theory versus experiment.}.

The key point is that, in the SM, CP-violating effects are absent.
Thus, the observation of a nonzero entry in Table \ref{tab:NI2} would
be a smoking-gun signal of NP.

\subsection{CP-conserving angular terms}

NP signals are not as easy to obtain from the measurement of Table
\ref{tab:NSR}. In each of the nine entries, the coefficient contains
at least one term involving the helicity amplitudes $\cA_0$,
$\cA_\parallel$, $\cA_\bot$ and $\cA_t$, all of which are present in
the SM. That is, even if there is no NP, all the angular functions of
Table \ref{tab:NSR} will be found in the angular distribution.

On the other hand, in the presence of NP, the coefficients will be
modified from their SM predictions. Thus, the way to detect NP is to
measure the coefficients in as many $q^2$-$E_\pi$ bins as possible,
and then perform a combined fit to all measurements and extract the
best-fit values of the magnitudes and relative weak phases of $1 +
g_L$, $g_R$, $g_P$ and $g_T$.

If a smoking-gun signal of NP has already been observed in the
measurement of Table \ref{tab:NI2}, the values of the NP parameters
responsible for it can be determined in this way. And even if no such
signal has been seen, the presence of CP-conserving NP can be detected
through the measurement of the angular distribution of Table
\ref{tab:NSR} in a sufficient number of different $q^2$-$E_\pi$ bins.

\section{Integrated Observables}

The full differential decay rate for $B\to D^*(\to
D\pi')\tau(\to\pi\nuta) \nutab$ depends on the five kinematic
parameters $q^2$, $E_\pi$, $\theta^*$, $\theta_\pi$ and $\chi_\pi$.
While a complete study of the decay distribution as a function of all
five parameters can reveal NP effects, a full experimental analysis
may be statistics limited. Effects of NP can still be studied through
``integrated observables,'' obtained by integrating the differential
decay rate over one or more of the kinematic parameters.

We separate the integrated observables into two types. The first type
is found by integrating over all three of the lepton-side parameters
($E_\pi$, $\theta_\pi$, $\chi_\pi$). Such observables are functions of
$q^2$, and are independent of the dynamics of the lepton decay. They
can, therefore, be used to study lepton-flavor
universality. Observables such as the longitudinal and transverse
$D^*$ polarizations ($F^{D^*}_{L,T}$) fall in this category. The
second type of observables are constructed by integrating over the
hadron-side parameter, $\theta^*$, and either of the parameters
$\theta_\pi$, and $\chi_\pi$. These observables explicitly depend on
the effects from the $\tau\to\pi\nuta$ decay.  Since lighter leptons
cannot decay to a pion, this second type of observables appears only
when the intermediate lepton is a $\tau$.

\subsection{Lepton flavor universality}

Here we consider observables constructed from the differential decay
distribution by integrating over $E_\pi$, $\theta_\pi$, $\chi_\pi$.
The resulting distribution in $q^2$ and $\theta^*$ can be expressed as
\bea
\frac{d^2\Ga}{dq^2 \, d\cos\theta^*} &=& \frac{3}{2} \, \frac{d\Ga}
{dq^2} \, \frac{a(q^2) + c(q^2)\cos^2\theta^*}{3a(q^2)+c(q^2)} \nl
&= & \frac{3}{4} \, \frac{d\Ga}
{dq^2} \left[ 2F^{D*}_L(q^2)\cos^2\theta^* + F^{D*}_T(q^2)\sin^2\theta^*\right] ~,
\label{eq:Plong}
\eea
where $F^{D*}_L(q^2)$ and $F^{D*}_T(q^2)= 1-F_L^{D*}(q^2)$ are the
longitudinal and transverse polarization fractions of the $D^*$. The
functions $a(q^2)$ and $c(q^2)$ are given by
\bea
a(q^2) &=& 2\(1 + \frac{m_\tau^2}{2q^2}\)\(|\cA_\parallel|^2 + |\cA_\bot|^2\)
+ 16 \( 1 + \frac{2m_\tau^2}{q^2}\)\(|\cA_{\parallel,T}|^2 + |\cA_{\bot,T}|^2\) \nl
       && \qquad -~\frac{24m_\tau}{\sqrt{q^2}} \(\Re[\cA_\parallel \cA_{\parallel,T}^{*}]
       + \Re[\cA_\bot\cA_{\bot,T}^*]\) ~, \\
c(q^2) &=& 2\(1 + \frac{m_\tau^2}{2q^2}\)\(2|\cA_0|^2 - |\cA_\parallel|^2 - |\cA_
\bot|^2\) + 6\lp\frac{m_\tau}{\sqrt{q^2}}\cA_t + \cA_{SP}\rp^2  \nl
        && \qquad +~16\(1 + \frac{2m_\tau^2}{q^2}\) \(2|\cA_{0,T}|^2 - |\cA_{\parallel,T}|^2
         - |\cA_{\bot,T}|^2\) \nl
        && \qquad -~\frac{24m_\tau}{\sqrt{q^2}}\(2\Re\[\cA_0 \cA_{0,T}^*\] -
        \Re[\cA_\parallel\cA_{\parallel,T}^*] - \Re[\cA_\bot \cA_{\bot,T}^*]\) ~.
\eea
The longitudinal and transverse polarization fractions $F^{D^*}_{L,T}$
can be obtained from Eq.~(\ref{eq:Plong}):
\beq
F^{D^*}_L = \frac{a(q^2)+c(q^2)}{3a(q^2) + c(q^2)} ~~,~~~~
F^{D^*}_T = \frac{2a(q^2)}{3a(q^2) + c(q^2)} ~.
\eeq
Further integration over $\cos\theta^*$ gives us the decay
distribution as a function of $q^2$:
\beq
\frac{d\Ga}{dq^2} ~=~ \frac{G_F^2|V_{cb}|^2 |\vp_{D^*}| q^2}{128 m_B^2 \pi^3}\,
\(1 - \frac{m^2_\tau}{q^2}\)^2\,\cB(D^*\to D\pi')\cB(\tau\to\pi\nuta)\(a(q^2) +
\frac{c(q^2)}{3}\) ~.
\label{q2dist}
\eeq

The integrated observables constructed above are not affected by the
dynamics of the $\tau$ decay, since the relevant kinematic parameters
have been integrated over. Indeed, the expressions for these
observables agree with those found elsewhere in the literature (apart
from the factor $\cB(\tau\to\pi\nuta)$ in Eq.~(\ref{q2dist})). The
comparison of the measured values of these observables with those
found in decays involving the light leptons, taking into account the
larger $\tau$ mass and the associated kinematic differences, provides
a test of lepton flavor universality.

\subsection{Lepton-side observables}

Here we discuss observables obtained by integrating the full differential
distribution over $\theta^*$ and either (or both) of $\theta_\pi$ and
$\chi_\pi$. These observables depend on at least one kinematic parameter
associated with the decay of the $\tau$, $E_\pi$. Therefore, these observables
can only be constructed in the $\tau$ lepton case, and specifically for the
decay $\tau\to\pi\nuta$.

The first step is to integrate the full differential decay rate of Eq.\
(\ref{eq:angdist}) over $\theta^*$. The $\theta^*$ dependence of the full
angular distribution can be retrieved from Tables \ref{tab:NSR} and
\ref{tab:NI2}. The angular functions in these Tables are proportional to
one of three forms -- $\cos^2\theta^*, \sin^2\theta^*$ and $\sin2\theta^*$.
The integral over $\theta^*$ eliminates all helicity-amplitude combinations
proportional to $\sin2\theta^*$, but keeps the other two. Thus, terms in the
angular distribution proportional to $f^R_{1,\ldots,7}$ and $f^I_3$ survive.
The remaining expression is long and may not carry any more insight than
the full angular distribution itself. We therefore proceed one step further
and integrate over $\chi_\pi$.

Once again, the $\chi_\pi$ dependence can be retrieved from Tables \ref{tab:NSR}
and \ref{tab:NI2}. Only terms that are independent of $\chi_\pi$ at this stage
still survive after we integrate over $\chi_\pi$. These terms appear in Table
\ref{tab:NSR} as those proportional to $f^R_{1,\ldots,6}$. The remaining
differential decay rate is a function of $q^2$, $E_\pi$ and $\theta_\pi$, and
can be expressed in terms of the functions $f_{1},\ldots,6$ as
\beq
\frac{d^3\Ga}{dq^2dE_\pi d\cos\theta_\pi} ~=~ \frac{3}{2}\,\frac{d^2\Ga}{dq^2dE_\pi}\,
\frac{a_\pi + b_\pi\cos\theta_\pi + c_\pi\cos^2\theta_\pi}{3a_\pi + c_\pi} ~,
\eeq
where the coefficients $a_\pi$, $b_\pi$ and $c_\pi$ are functions of
$q^2$ and $E_\pi$ (the $f^R_i(q^2,E_\pi)$ are defined in Table
\ref{tab:NSR}):
\bea
a_\pi &=& f^R_1(q^2,E_\pi) + 2f^R_2(q^2,E_\pi) - f^R_4(q^2,E_\pi) - 2f^R_6(q^2,E_\pi) \nn\\
&=& (S_{0,1}-S_{0,2})|\cA_0 |^2 + (S_{0T,1} - S_{0T,2})|\cA_{0,T}|^2 + S_{SP}|\cA_{SP}|^2
      + S_t |\cA_t|^2 \nl
      && +~ 2 (S_{\parallel,1} - S_{\parallel,2}) |\cA_{\parallel }|^2 + 2(S_{\parallel T,1}
      -  S_{\parallel T,2}) |\cA_{\parallel, T}|^2 + 2( S_{\bot ,1} -  S_{\bot ,2}) |\cA_{\bot}|^2 ~ \nl
      && +~ 2 (S_{\bot T,1} -  S_{\bot T,2}) |\cA_{\bot ,T}|^2 + (R_{0T0,1} - R_{0T0,2})  \, \Re[\cA_{0,T}\cA_0^*]
       + R_{SPt}  \, \Re[\cA_{SP}\cA_t^*] \nl
      && +~ 2( R_{\parallel T\parallel ,1}-  R_{\parallel T\parallel ,2})  \, \Re[\cA_{\parallel ,T}\cA_{\parallel }^*]
        + 2( R_{\bot T\bot ,1} -  R_{\bot T\bot ,2})  \, \Re[\cA_{\bot ,T}\cA_{\bot }^*] ~, \\
&& \nl
\label{bpi}
b_\pi &=& f^R_3(q^2,E_\pi) + 2f^R_5(q^2,E_\pi) \nn\\
&=& ~ R_{0Tt}  \, \Re[\cA_{0,T}\cA_t^*]  +  R_{SP0}  \, \Re[\cA_{SP}\cA_0^*]  + R_{SP0T}  \, \Re[\cA_{SP}\cA_{0, T}^*]
      + R_{t0}  \, \Re[\cA_t \cA_0^*]  \nl
      && +~ 2 R_{\parallel T\bot }  \, \Re[\cA_{\parallel,T}\cA_{\bot }^*]  + 2 R_{\parallel \bot}  \, \Re[\cA_{\parallel}
      \cA_{\bot }^*]  + 2 R_{\bot T\parallel}  \, \Re[\cA_{\bot ,T}\cA_{\parallel }^*] ~ \nl
      && +~ 2 R_{\bot T\parallel T}  \, \Re[\cA_{\bot ,T}\cA_{\parallel ,T}^*]  ~, \\
&& \nl
c_\pi &=& 2 f^R_4(q^2,E_\pi) + 4f^R_6(q^2,E_\pi) \nn\\
&=& ~ 2 S_{0,2} |\cA_0|^2 + 2 S_{0T,2} |\cA_{0,T}|^2 + 4 S_{\parallel,2}
      |\cA_{\parallel }|^2  \nl
      && +~ 4 S_{\parallel T,2} |\cA_{\parallel ,T}|^2 + 4 S_{\bot ,2} |\cA_{\bot }|^2 + 4 S_{\bot T,2}
      |\cA_{\bot ,T}|^2 \nl
      && +~ 2 R_{0\text{T0},2}  \, \Re[\cA_{0,T}\cA_0^*]  + 4 R_{\parallel T\parallel ,2}  \, \Re[\cA_{\parallel ,T}\cA_
      {\parallel}^*]  + 4 R_{\bot T\bot ,2}  \, \Re[\cA_{\bot ,T}\cA_{\bot }^*]   ~.
\eea

Further integrating over $\theta_\pi$ gives us the decay distribution
as a function of $q^2$ and $E_\pi$:
\beq
\frac{d^2\Ga}{dq^2dE_\pi} ~=~ \frac{m_\tau^2\,(q^2)^{5/2}}{2\,(m_\tau^2-m_\pi^2)^2\,(q^2 - m^2_\tau)
^2}\,\frac{3\,a_\pi + c_\pi}{3\,a(q^2) + c(q^2)} \, \frac{d\Ga}{dq^2} ~.
\eeq

At this stage, we can perform an asymmetric integral over
$\cos\theta_\pi$, to find the forward-backward asymmetry in the
distribution of the $\pi$ coming from the $\tau$ decay. This is done
by integrating the differential decay rate with a uniform negative
weight for the positive values of $\cos\theta_\pi$, and subtracting
this from a similar integral with a uniform positive weight for the
negative values of $\cos\theta_\pi$. Appropriately normalizing this
function, we can define the forward-backward asymmetry ($A_{FB}$) as
follows:
\bea
A_{FB}(q^2,E_\pi) &\equiv& \frac{\int\limits_{-1}^0 \frac{d^3\Ga}{dq^2dE_\pi d\cos\theta_\pi}
d\cos\theta_\pi - \int\limits_{0}^1 \frac{d^3\Ga}{dq^2dE_\pi d\cos\theta_\pi}d\cos\theta_\pi}
{\frac{d^2\Ga}{dq^2dE_\pi}} ~, \nl
&=& -\frac{3}{2}\,\frac{b_\pi}{3\,a_\pi + c_\pi} ~.
\eea
As can be seen from the form of $b_\pi$ [Eq.~(\ref{bpi})], $A_{FB}$ is
nonzero in the SM. In order to see if NP is present, one must combine
this measurement with that of other observables, or of other terms in
the angular distribution. With enough independent measurements of
functions of the helicity amplitudes, it is possible to determine if
some NP amplitudes must be nonzero.

Changing the order of integrals over $\chi_\pi$ and $\theta_\pi$ can
yield valuable complementary information. In the preceding discussion
we obtained observables by first integrating over $\chi_\pi$ and then
over $\theta_\pi$. If instead the integral over $\theta_\pi$ is performed
first, the helicity-amplitude combinations proportional to $\cos\theta_\pi$
and $\sin\theta_\pi$ are removed. The terms in the angular distribution
proportional to $f^R_{1,2,4,6,7}$ and $f^I_3$ survive. The remaining
differential decay rate, a function of $q^2$, $E_\pi$ and $\theta_\pi$,
is found to be
\beq
\frac{d^3\Gamma}{dq^2dE_\pi d\chi_\pi} ~=~ \frac{d^2\Ga}{dq^2dE_\pi}\, \frac{B_1 + B_2
\cos2\chi_\pi + B_3\sin 2\chi_\pi}{2\pi B_1} ~,~~
\eeq
where the coefficients $B_i$ are functions of $q^2$ and $E_\pi$ (the
$f^R_i(q^2,E_\pi)$ are defined in Table \ref{tab:NSR}):
\bea
B_1 &=& 3f^R_1(q^2,E_\pi) + 6f^R_2(q^2,E_\pi) - f^R_4(q^2,E_\pi) - 2f^R_6(q^2,E_\pi)  \nn\\
&=& (3S_{0,1}-S_{0,2})|\cA_0|^2 + (3S_{0T,1}-S_{0T,2})|\cA_{0,T}|^2 + 3S_{SP}|\cA_{SP}|^2
       + 3S_t|\cA_t|^2  \nl
    && + 2(3S_{\parallel,1}-S_{\parallel,2})|\cA_{\parallel }|^2 + 2(3S_{\parallel T,1}-S_{\parallel T,2}) |\cA_{\parallel ,T}|^2 + 2(3S_{\bot ,1}-S_{\bot ,2}) |\cA_{\bot }|^2   \nl
         &&  + 2(3S_{\bot T,1}-S_{\bot T,2})|\cA_{\bot ,T}|^2  +  (3R_{0T0,1}-R_{0T0,2}){\rm Re}[\cA_{0,T}\cA_0^*]
         + 3 R_{SPt} {\rm Re}[\cA_{SP}\cA_t^* ]   \nl
      &&   + 2(3R_{\parallel T\parallel ,1}-R_{\parallel T\parallel ,2}) {\rm Re}[\cA_{\parallel ,T}\cA_{\parallel }^*]
           + 2 (3R_{\bot T\bot ,1}-R_{\bot T\bot ,2}){\rm Re}[\cA_{\bot ,T}\cA_{\bot }^* ]~, \\
&& \nl
B_2 &=& 4 f^R_7(q^2,E_\pi) \nn\\
&=& 8\(S_{\bot,2}\lp\cA_{\bot }\rp^2 - S_{\parallel,2} \lp\cA_{\parallel}\rp^2\) + 8\(S_{\bot T,2}\lp\cA_{\bot ,T}\rp^2 -  S_{\parallel T,2}\lp\cA_{\parallel ,T}\rp^2\) \nn\\
&&+~8 \(R_{\bot T \bot,2} \, \Re[\cA_{\bot,T}\cA_{\bot}^*] - R_{\parallel T \parallel,2} \, \Re[\cA_{\parallel,T}\cA_{\parallel}^*]\)~,  \\
&& \nl
\label{B3}
B_3 &=& 4f^I_3(q^2,E_\pi) \nn\\
&=& 4 \(I_{\parallel \bot} \Im[\cA_{\parallel}\cA_{\bot}^*] + I_{\parallel T\bot} \Im[\cA_{\parallel ,T} \cA_{\bot }^*]
    + I_{\bot T\parallel } {\rm Im}[\cA_{\bot ,T} \cA_{\parallel }^* ]\) ~.
\eea
Note that the coefficient $B_1$ is related to the coefficients
$a_\pi$ and $c_\pi$:
\beq
B_1 ~=~ 3\,a_\pi + c_\pi ~.
\eeq

An asymmetric integral over $\chi_\pi$ can now isolate an observable
that is nonzero only if true CP-violating TP asymmetries are
present. This new observable, $A_{TP}$, can be defined as
\beq
A_{TP}(q^2,E_\pi) ~=~ \(\frac{d^2\Ga}{dq^2dE_\pi}\)^{-1}\(\int\limits_{0}^{\pi/2}
- \int\limits_{\pi/2}^{\pi} + \int\limits_{\pi}^{3\pi/2} - \int\limits_{3\pi/2}^{2\pi}\)
\frac{d^3\Ga}{dq^2dE_\pi d\chi_\pi} d\chi_\pi ~=~ \frac{B_3}{2\pi B_1} ~.
\eeq
From Eq.~(\ref{B3}), we see that $B_3$ vanishes in the absence of
weak-phase differences. This shows that $A_{TP}$ is a CP-violating
observable.

Above, $A_{TP}$ is defined as a function of $q^2$ and $E_\pi$.
However, one can further integrate this function over both of these
variables.  The resulting integrated observable can be directly
compared to an experimental event analysis in which one obtains the
asymmetry between the number of events with $\sin2\chi_\pi > 0$ and
$\sin2\chi_\pi < 0$.

We note that $A_{TP}$ involves interferences of vector-vector and
vector-tensor type. The only way to generate a nonzero value of
$A_{TP}$ is if there is a nonzero weak phase in at least one of $g_R$
and $g_T$.  Thus, if $A_{TP}$ (and/or its form integrated over $q^2$
and $E_\pi$) is found to be nonzero, this will be an unmistakeable
sign of CP-violating NP.

\section{Conclusions}

At the present time, the measurements of $R_{D^{(*)}} \equiv
\cB(\bar{B} \to D^{(*)} \tau^{-} {\bar\nu}_\tau)/\cB(\bar{B} \to
D^{(*)} \ell^{-} {\bar\nu}_\ell)$ ($\ell = e,\mu$) and $R_{J/\psi}
\equiv \cB(B_c^+ \to J/\psi\tau^+\nu_\tau) / \cB(B_c^+ \to
J/\psi\mu^+\nu_\mu)$ disagree with the predictions of the SM, hinting
at NP in $\bctaunu$ decays. There are many possibilities for this
NP. A variety of observables have been proposed to distinguish the
various NP explanations: the $q^2$ distribution, the $D^*$
polarization, the $\tau$ polarization, etc.

Another potential way of distinguishing the NP explanations involves
CP violation.  Within the SM, there are no CP-violating effects in
${\bar B} \to D^{*} \tau^- {\bar\nu}_\tau$, so that any observation of
CP violation in this decay would be a smoking-gun signal of NP. Here,
the main CP-violating effects appear as CP-violating asymmetries in
the angular distribution. However, this is problematic.  The
construction of the angular distribution requires the knowledge of the
three-momentum ${\vec p}_\tau$. But this cannot be measured precisely,
since the $\tau$ decays to final-state particles that include
$\nu_\tau$, which is undetected. The result is that the full angular
distribution in ${\bar B} \to D^* (\to D \pi) \, \tau^{-}
{\bar\nu}_\tau$ cannot be measured.

In this paper, we construct a {\it measurable} angular distribution by
considering the additional decay $\tau^- \to \pi^- \nu_\tau$. The full
process then is ${\bar B} \to D^* (\to D \pi') \, \tau^{-} (\to \pi^-
\nu_\tau) {\bar\nu}_\tau$. Here there are three final-state particles
whose three-momenta {\it can} be measured: the $D$ and $\pi'$ (from
$D^*$ decay), and the $\pi^-$ (from $\tau$ decay). The new angular
distribution is given in terms of five kinematic parameters: $q^2$,
$\theta^*$ (describing $D^* \to D\pi$), and three quantities
describing the $\pi^-$, $E_\pi$, $\theta_\pi$ and $\chi_\pi$. It
includes CP-violating angular asymmetries, which can be measured and
used to extract information about the NP.

But much more information can be extracted from the angular
distribution. In the most general case, the angular distribution
involves the couplings $1 + g_L$, $g_R$, $g_P$ and $g_T$, where $g_L$,
$g_R$, $g_P$ and $g_T$ are the NP parameters. The magnitudes and
relative phases of all four couplings can be extracted from a fit to
the full distribution. This will go a long way towards identifying the
NP.

It is also possible to integrate over one or more of the five
kinematic parameters. If one integrates over the lepton-side
parameters $E_\pi$, $\theta_\pi$ and $\chi_\pi$, all the familiar
observables that have been proposed to distinguish NP models are
reproduced. These include the $q^2$ distribution and the $D^*$
polarization. And if one integrates over the hadron-side quantities,
one obtains new observables that depend on the kinematic angles
associated with the $\pi^-$ emitted in the $\tau$ decay, $\theta_\pi$
and $\chi_\pi$. These include the forward-backward asymmetry of the
$\pi^-$, and the CP-violating triple-product asymmetry.

In principle, one can construct angular distributions using other
$\tau$ decays.  The analysis of $\tau \to 3\pi\nu_\tau$ is similar to
that of $\tau \to \pi\nu_\tau$, treating the $3\pi$ system as a single
``particle,'' except that one must allow for it to have spin 0, 1, 2,
etc. And $\tau \to \mu {\bar\nu}_\mu \nu_\tau$ is more complicated,
since one must also integrate over the kinematic angles of the
${\bar\nu}_\mu$.

\bigskip
{\bf Acknowledgments}: This work was financially supported in part by
the NSF Grant No. PHY-1915142 (AD \& SK), and NSERC of Canada
(DL). Initial work of BB was supported by Lawrence Technological
University's Faculty Seed Grant. BB thanks D.~Cinabro,
W.~Altmannshofer, and R.~Zwicky for useful discussions.

\appendix

\section{Hadronic and Helicity Amplitude Pieces}
\label{app:helicity}

The differential decay rate for the process $B\to D^* (\to D\pi') \tau
(\to \pi \nuta) \nutab$ has been written in terms of a collection of
hadronic pieces $\cH_{D^*}$, helicity amplitude pieces $\cM^{N^*}$,
and leptonic pieces $\tL^{N^*}$ in Section \ref{sec:msq}. While the
leptonic pieces are new in this analysis, the hadronic and helicity
amplitudes were presented in Ref.\ \cite{BDmunuCPV}. For
convenience, below we summarize these relationships.

The hadronic pieces, $H_{D^*}$, can be expressed as
\beq
\cH_{D^*}(m) ~=~ \ep_{D^*}(m)\cdot p_D ~~,~~~~ m ~=~ 0,\pm ~,
\eeq
where $p_D$ represents the four-momentum of the $D$ meson, and
$\ep_{D^*}(m)$ represents the polarization of the $D^*$ meson. We
follow the convention of expressing the four-momentum and
polarizations of the $D^*$ meson in the $B$-meson rest frame as
follows:
\beq
p_{D^*}^\mu ~=~ (k_0,0,0,k_z) ~~,~~~~ \ep^\mu_{D^*}(\pm) ~=~ (0,1,\pm{\it i},0)/\s
~~,~~~~ \ep^\mu_{D^*}(0) ~=~ (k_z,0,0,k_0)/m_{D^*} ~.~~~~
\eeq
In addition to the hadronic pieces corresponding to the three
well-defined helicities of the on-shell $D^*$ meson, we make use of a
fourth timelike helicity for an off-shell particle, defined such that
$\cH_{D^*}(t) = \cH_{D^*}(0)$.

The helicity amplitude pieces, $\cM^{N^*}$, also depend on the
helicities of the intermediate particles. These are of the
scalar-pseudoscalar ($SP$), the vector-axialvector ($VA$), and the
tensor ($T$) types. Furthermore, since the decaying $B$ meson is
spinless, only certain helicity combinations survive. The list of
non-zero components of the helicity amplitude pieces are listed below:
\bea
& \cM^{SP}_{(0)}(B\to D^* SP^*) = \cA_{SP} ~,~~ & \nl
& \cM^{VA}_{(+;+)}(B\to D^* VA^*) = \cA_+ ~,~~ & \nl
& \cM^{VA}_{(-;-)}(B\to D^* VA^*) = \cA_- ~,~~ & \nl
& \cM^{VA}_{(0;0)}(B\to D^* VA^*) = \cA_0 ~,~~ & \nl
& \cM^{VA}_{(0;t)}(B\to D^* VA^*) = \cA_t ~,~~ & \nl
& \cM^T_{(+;+,0)}(B\to D^*T^*) = \cM^T_{(+;+,t)}(B\to D^*T^*) ~=~ \cA_{+,T} ~,~~ & \nl
& \cM^T_{(0;-,+)}(B\to D^*T^*) = \cM^T_{(0;0,t)}(B\to D^*T^*) ~=~ \cA_{0,T} ~,~~ & \nl
& \cM^T_{(-;0,-)}(B\to D^*T^*) = \cM^T_{(-;-,t)}(B\to D^*T^*) ~=~ \cA_{-,T} ~.~~ &
\label{eq:Tampdef}
\eea

As seen in the above, there are a total of 8 independent helicity
amplitudes: one of type $SP$, four of type $VA$, and three independent
amplitudes of type $T$. Using the definitions for the $B\to D^*$ form
factors $A_i(q^2)$, $V(q^2)$ and $T_i(q^2)$ given in
Refs.\ \cite{Sakaki:2013bfa, Beneke:2000wa}, we can further represent
each helicity amplitude as follows:
\bea
\cA_{SP} &=& -g_P \, \frac{\sqrt{\la(m_B^2,\mDs^2, q^2)}}{m_b+m_c} A_0(q^2) ~, \nl
\cA_{0} &=& -\frac{(1+g_L-g_R)\,(m_B + \mDs)}{2\mDs \sqrt{q^2}} \, \((m_B^2-\mDs^2 -q^2) A_1(q^2)
+ \frac{\la(m_B^2, \mDs^2 , q^2)}{(m_B + \mDs)^2} A_2(q^2)\) ~, \nl
\cA_{t} &=& -(1+g_L-g_R) \, \frac{\sqrt{\la(m_B^2,\mDs^2, q^2)}}{\sqrt{q^2}} A_0(q^2) ~, \nl
\cA_{\pm} &=& (1+g_L-g_R) \, (m_B + \mDs)A_1(q^2) \mp (1+g_L+g_R)\frac{\sqrt{\la(m_B^2,\mDs^2, q^2)}}
{m_B + \mDs} V(q^2)~, \nl
\cA_{0, T} &=& \frac{g_{T}}{2\mDs} \((m_B^2+3\mDs^2 -q^2)T_2(q^2)
- \frac{\la(m_B^2,\mDs^2, q^2)T_3(q^2)}{m_B^2-\mDs^2} \) ~, \nl
\cA_{\pm, T} &=&  g_{T} \, \frac{\sqrt{\la(m_B^2,\mDs^2, q^2)}T_1(q^2) \pm (m_B^2-\mDs^2)T_2(q^2)}
{\sqrt{q^2}} ~,
\eea
where $\la(a,b,c) = a^2 + b^2 + c^2 - 2 a b - 2 b c - 2 c a$~.

Finally, the amplitudes for the vector and tensor types can be expressed in the transversity
basis (using $\bot,\parallel$) instead of the helicity basis (using $\pm$), using the following
relationships,
\bea
\cA_{\parallel(,T)} &=& (\cA_{+(,T)} + \cA_{-(,T)})/\s ~, \nl
\cA_{\bot(,T)} &=& (\cA_{+(,T)} - \cA_{-(,T)})/\s ~.
\eea

\section{\boldmath $\cN_i^S$, $\cN_{i,j}^R$ and $\cN_{i,j}^I$ Contributions}
\label{app:functions}

the information relevant for the $\cN_i^S \, |\cA_i|^2$, $\cN_{i,j}^R
\, {\rm Re}[\cA_i\cA_j^*]$ and $\cN_{i,j}^I \, {\rm Im}[\cA_i
  \cA_j^*]$ pieces of Eq.~(\ref{eq:angdist}) are found in Tables
\ref{tab:NS}, \ref{tab:NR} and \ref{tab:NI}, respectively. The
dependence on $q^2$ and $E_pi$ is contained in the $S_i, R_i$, and
$I_i$ functions, respectively, whose expressions are given below.

\begin{table}
\begin{center}
\begin{tabular}{|c|c|} \hline
Helicity info & $\cN^S$ \\ \hline
$\lp\cA_t\rp^2$ & $S_t\,\cos^2 \theta^*$ \\ \hline
$\lp\cA_0\rp^2$ & $\left[ S_{0,1} + S_{0,2} \, \cos 2 \theta_\pi \right] \cos^2 \theta^*$ \\ \hline
$\lp\cA_{\bot }\rp^2$ & $\left[ S_{\bot ,1} + S_{\bot ,2} \left( \cos 2 \chi_\pi + 2 \cos 2 \theta_\pi \sin^2 \chi_\pi \right) \right] \sin^2 \theta^*$ \\ \hline
$\lp\cA_{\parallel }\rp^2$ & $ \left[ S_{\parallel ,1}  + S_{\parallel ,2} \left( \cos2\theta_\pi -2 \sin^2\theta_\pi \cos2\chi_\pi \right) \right] \sin^2\theta^*$ \\  \hline
$\lp\cA_{SP}\rp^2$ & $S_{SP} \cos^2 \theta^*$ \\ \hline
$\lp\cA_{0,T}\rp^2$ & $\left[ S_{0T,1} + S_{0T,2} \cos 2 \theta_\pi \right] \cos^2 \theta^*$ \\ \hline
$\lp\cA_{\bot ,T}\rp^2$ & $\left[ S_{\bot T,1} + S_{\bot T,2} \left( \cos 2 \theta_{\pi } + 2 \cos 2 \chi_\pi
   \sin^2 \theta_\pi \right) \right] \sin^2 \theta^*$ \\ \hline
$\lp\cA_{\parallel ,T}\rp^2$ & $\left[ S_{\parallel T,1} + S_{\parallel T,2} \left( \cos 2 \theta_\pi - 2 \cos 2 \chi_\pi \sin^2 \theta_\pi \right)\right] \sin^2 \theta^*$ \\ \hline
\end{tabular}
\caption{Contributions to the $\cN_i^S \, |\cA_i|^2$ pieces of
  Eq.~(\ref{eq:angdist}). The coefficients $S_i$ depend on the
  kinematic parameters $q^2$ and $E_\pi$, and are listed in
  Eq.~(\ref{Sifactors}).
\label{tab:NS}}
\end{center}
\end{table}

\begin{table}
\begin{center}
\begin{tabular}{|c|c|} \hline
Helicity info & $\cN^R$ \\ \hline
$\Re[\cA_t\cA_0^*]$ & $R_{t0} \,  \cos\theta_\pi \cos^2\theta^*$ \\ \hline
$\Re[\cA_t\cA_{\parallel }^*]$ & $R_{t\parallel } \,
\cos\chi_\pi \sin\theta_\pi \sin 2\theta^*$ \\  \hline
$\Re[\cA_0\cA_{\parallel }^*]$ & $R_{0\parallel } \, \cos\chi_\pi \sin 2\theta_\pi \sin 2\theta^*$ \\ \hline
$\Re[\cA_0\cA_{\bot }^*]$ & $R_{0\bot } \, \cos\chi_\pi \sin\theta_\pi \sin 2\theta^*$ \\ \hline
$\Re[\cA_{\parallel }\cA_{\bot }^*]$ & $R_{\parallel \bot } \,  \cos\theta_\pi \sin^2\theta^*$ \\ \hline
$\Re[\cA_{SP}\cA_t^*]$ & $R_{SPt} \,  \cos^2\theta^*$ \\ \hline
$\Re[\cA_{SP}\cA_0^*]$ & $R_{SP0} \,  \cos\theta_\pi \cos^2\theta^*$ \\ \hline
$\Re[\cA_{SP}\cA_{\parallel}^*]$ & $R_{SP\parallel} \,  \cos\chi_\pi \sin\theta_\pi \sin 2\theta^*$ \\ \hline
$\Re[\cA_{SP}\cA_{0,T}^*]$ & $R_{SP0T} \,  \cos\theta_\pi \cos^2\theta^*$ \\ \hline
$\Re[\cA_{SP}\cA_{\parallel ,T}^*]$ & $  R_{SP \parallel T }  \,  \cos\chi_\pi \sin\theta_\pi \sin 2\theta^*$ \\  \hline
$\Re[\cA_{0,T}\cA_t^*]$ & $R_{0Tt} \,  \cos\theta_\pi \cos^2\theta^*$ \\  \hline
$\Re[\cA_{0,T}\cA_0^*]$ & $\left[R_{0T0,1} +R_{0T0,2} \,  \cos 2\theta_\pi\right] \cos^2\theta^*$ \\  \hline
$\Re[\cA_{0,T}\cA_{\bot }^*]$ & $R_{0T\bot }  \, \cos\chi_\pi \sin\theta_\pi \sin 2\theta^*$ \\  \hline
$\Re[\cA_{0,T}\cA_{\parallel }^*]$ & $R_{0T\parallel}  \, \cos\chi_\pi \sin 2\theta_\pi \sin 2\theta^*$ \\  \hline
$\Re[\cA_{0,T}\cA_{\bot ,T}^*]$ & $R_{0T\bot T} \,  \cos\chi_\pi  \sin\theta_\pi \sin 2\theta^*$ \\  \hline
$\Re[\cA_{0,T}\cA_{\parallel ,T}^*]$ & $R_{0T\parallel T} \,  \cos\chi_\pi
   \sin 2\theta_\pi \sin 2\theta^*$ \\  \hline
$\Re[\cA_{\bot ,T}\cA_0^*]$ & $R_{\bot T0} \, \cos\chi_\pi \sin\theta_\pi \sin 2\theta^*$ \\  \hline
$\Re[\cA_{\bot ,T}\cA_{\bot }^*]$ & $\left[R_{\bot T\bot ,1} + R_{\bot T\bot ,2}  \left(\cos 2\theta_\pi \right. \right.$ \\
& $\left. \left. \hskip2truecm +~2 \cos 2 \chi_\pi \sin^2\theta_\pi\right)\right]
   \sin^2\theta^*$ \\  \hline
$\Re[\cA_{\bot ,T}\cA_{\parallel }^*]$ & $R_{\bot T\parallel } \,   \cos\theta_\pi \sin^2\theta^*$ \\  \hline
$\Re[\cA_{\bot ,T}\cA_{\parallel ,T}^*]$ & $R_{\bot T\parallel T} \,  \cos\theta_\pi \sin^2\theta^*$ \\  \hline
$\Re[\cA_{\parallel ,T}\cA_t^*]$ & $R_{\parallel Tt} \,  \cos\chi_\pi \sin\theta_\pi \sin 2\theta^*$ \\ \hline
$\Re[\cA_{\parallel ,T}\cA_0^*]$ & $R_{\parallel T0}  \, \cos\chi_\pi \sin 2\theta_\pi \sin 2\theta^*$ \\  \hline
$\Re[\cA_{\parallel ,T}\cA_{\bot }^*]$ & $R_{\parallel T\bot } \,  \cos\theta_\pi \sin^2\theta^*$ \\  \hline
$\Re[\cA_{\parallel ,T}\cA_{\parallel }^*]$ & $\left[R_{\parallel T\parallel ,1} +R_{\parallel T\parallel ,2}  \left(\cos 2\theta_\pi \right. \right.$ \\
& $\left. \left. \hskip2truecm -~2 \cos 2 \chi_\pi \sin^2\theta_\pi\right)\right] \sin^2\theta^*$ \\ \hline
\end{tabular}
\caption{Contributions to the $\cN_{i,j}^R \, \Re[\cA_i\cA_j^*]$
  pieces of Eq.~(\ref{eq:angdist}). The coefficients $R_i$ depend on
  the kinematic parameters $q^2$ and $E_\pi$, and are listed in
  Eq.~(\ref{Rifactors}).
\label{tab:NR}}
\end{center}
\end{table}

\begin{table}
\begin{center}
\begin{tabular}{|c|c|} \hline
Helicity info & $\cN^I$ \\ \hline
$\Im[\cA_t \cA_{\bot}^*]$ & $I_{t\bot}        \sin2\theta^* \sin\theta_\pi \sin\chi_\pi$ \\ \hline
$\Im[\cA_{\parallel ,T} \cA_0^*]$ & $I_{\parallel T0} \sin2\theta^* \sin\theta_\pi \sin\chi_\pi$ \\
\hline
$\Im[\cA_{SP}\cA_{\bot}^*]$   & $I_{SP\bot}           \sin2\theta^* \sin\theta_\pi \sin\chi_\pi$ \\ \hline
$\Im[\cA_{SP}\cA_{\bot,T}^*]$ & $I_{SP\bot T}          \sin2\theta^* \sin\theta_\pi \sin\chi_\pi$ \\ \hline
$\Im[\cA_{0,T} \cA_{\parallel }^*]$ & $I_{0T\parallel}\sin2\theta^* \sin\theta_\pi \sin\chi_\pi$ \\
\hline
$\Im[\cA_{\bot ,T} \cA_t^*]$ & $I_{\bot Tt}           \sin2\theta^* \sin\theta_\pi \sin\chi_\pi$ \\ \hline
$\Im[\cA_0 \cA_{\bot}^*]$ & $I_{0\bot} \sin2\theta^* \sin2\theta_\pi \sin\chi_\pi$ \\ \hline
$\Im[\cA_{0,T} \cA_{\bot }^*]$ & $I_{0T\bot}   \sin2\theta^* \sin2\theta_\pi \sin\chi_\pi$ \\ \hline
$\Im[\cA_{\bot ,T} \cA_0^*]$ & $I_{\bot T0}    \sin2\theta^* \sin2\theta_\pi \sin\chi_\pi$ \\ \hline
$\Im[\cA_{\parallel}\cA_{\bot}^*]$ & $I_{\parallel \bot } \sin^2\theta^* \sin^2\theta_\pi \sin 2\chi_\pi$ \\  \hline
$\Im[\cA_{\bot ,T} \cA_{\parallel }^* ]$ & $I_{\bot T\parallel} \sin^2\theta^* \sin^2\theta_\pi \sin 2\chi_\pi$ \\  \hline
$\Im[\cA_{\parallel ,T} \cA_{\bot }^*]$ & $I_{\parallel T\bot} \sin^2\theta^* \sin^2\theta_\pi \sin 2\chi_\pi$ \\ \hline
\end{tabular}
\caption{Contributions to the $\cN_{i,j}^I \, \Im[\cA_i \cA_j^*]$
  pieces of Eq.~(\ref{eq:angdist}). The coefficients $I_i$ depend on
  the kinematic parameters $q^2$ and $E_\pi$, and are listed in
  Eq.~(\ref{Iifactors}).
\label{tab:NI}}
\end{center}
\end{table}

The kinematics of the five-body decay restricts the range of values that
the parameters $q^2$ and $E_\pi$ can take.
\beq
m^2_\tau \leq q^2 \leq (m_B - m_{D^*})^2 ~,~~~~\frac{m^4_\tau + m^2_\pi
q^2}{2m^2_\tau\sqrt{q^2}}\leq E_\pi \leq \frac{q^2 + m^2_\pi}{2\sqrt{q^
2}} ~.
\eeq
We define the normalized parameters $\rho_\tau\equiv m_\tau/\sqrt{q^2},
\rho_\pi\equiv m_\pi/\sqrt{q^2}$, and $\mathcal{E}_\pi\equiv E_\pi/
\sqrt{q^2}$. Based on the above limits, the normalized parameters are
limited to values between 0 and 1. Below we express $S_i, R_i$, and $I_i$
in terms of these normalized parameters.

The expressions for the $S_i$ factors are
\bea
\label{Sifactors}
S_t = \rho_\tau^2 S_{SP} &=& 16 \rho _{\tau }^2 \left(2 \mathcal{E}_{\pi } \rho _{\tau }^2-\rho _{\tau }^4-\rho_\pi^2\right)  ~, \nn\\
S_{0,1} &=& \frac{1}{\mathcal{E}_{\pi }^2-\rho_\pi^2}     \left(4 \rho _{\tau }^2\right)  \Big( 2 \mathcal{E}_{\pi } \left(\rho _{\tau }^2 \left(1-\rho_\pi^2\right)+\rho _{\tau }^4 + 3 \rho_\pi^2\right) \nn \\
       && \hskip3truecm
+~2 \mathcal{E}_{\pi }^2 \left(1-2 \rho _{\tau }^2-\rho _{\tau }^4\right) - 4 \mathcal{E}_{\pi }^3 \left(1-\rho _{\tau }^2\right) \nn\\
&& \hskip3truecm +~\rho _{\tau }^4 \left(-1+\rho_\pi^2\right) - 3 \rho_\pi^2-\rho_\pi^4 \Big) ~, \nn\\
S_{0,2} = -4 S_{\bot ,2} &=& \frac{ 1}{\mathcal{E}_{\pi }^2-\rho_\pi^2}      \left(4 \rho _{\tau }^2\right) \Big( 2 \mathcal{E}_{\pi } \left(1+\rho _{\tau }^2\right) \left(3 \rho _{\tau }^2+\rho_\pi^2\right)  \nn \\
     && \hskip3truecm
-~2 \mathcal{E}_{\pi }^2 \left(1+6 \rho _{\tau }^2+\rho _{\tau }^4 + 2 \rho_\pi^2\right)
+ 4 \mathcal{E}_{\pi }^3 \left(1+\rho _{\tau }^2\right) \nn\\
&& \hskip3truecm -~\rho _{\tau }^4 \left(3+\rho_\pi^2\right)-\rho_\pi^2+\rho_\pi^4 \Big) ~, \nn\\
S_{\bot ,1} &=& \frac{1}{\mathcal{E}_{\pi }^2-\rho_\pi^2}    \left(-\rho _{\tau }^2\right) \Big (2 \mathcal{E}_{\pi } \left(\rho _{\tau }^2 \left(1+3 \rho_\pi^2\right)+\rho _{\tau }^4-5 \rho_\pi^2\right) \nn \\
           && \hskip3truecm
-~2 \mathcal{E}_{\pi }^2 \left(3+2 \rho _{\tau }^2-\rho _{\tau }^4 + 2 \rho_\pi^2\right)
+ 4 \mathcal{E}_{\pi }^3 \left(3-\rho _{\tau }^2\right) \nn\\
&& \hskip3truecm -~\rho _{\tau }^4 \left(1+3 \rho_\pi^2\right)+5 \rho_\pi^2+3 \rho_\pi^4 \Big)  ~, \nn\\
S_{\parallel,1} &=& \frac{\rho_{\tau }^2}{\mathcal{E}_{\pi }^2-\rho _{\pi }^2} \left( -2 \mathcal{E}_{\pi } \left(\rho _{\tau }^2+\rho _{\tau }^4+\rho_{\pi }^2 \left(-5+3 \rho _{\tau }^2\right)\right) + \mathcal{E}_{\pi }^2 \left(6+4 \rho_{\pi }^2+4 \rho _{\tau }^2-2 \rho _{\tau }^4\right)+4 \mathcal{E}_{\pi }^3 \left(-3+\rho_{\tau }^2\right) \right. \nonumber \\
                && \left. -3 \rho _{\pi }^4+\rho _{\tau }^4+\rho _{\pi }^2 \left(-5+3 \rho _{\tau
   }^4\right)\right)  \nonumber \\
S_{\parallel,2} &=& \frac{\rho _{\tau }^2}{\mathcal{E}_{\pi }^2-\rho _{\pi }^2} \left(-2 \mathcal{E}_{\pi } \left(1+\rho _{\tau }^2\right) \left(\rho_{\pi }^2+3 \rho _{\tau }^2\right)+2 \mathcal{E}_{\pi }^2 \left(1+2 \rho _{\pi }^2+6 \rho_{\tau }^2+\rho _{\tau }^4\right) - 4 \mathcal{E}_{\pi }^3 \left(1+\rho _{\tau}^2\right) \right. \nonumber \\
                && \left. + \rho _{\pi }^2-\rho _{\pi }^4+\left(3+\rho _{\pi }^2\right) \rho _{\tau
   }^4\right) \nonumber \\
S_{0T,1} &=& \frac{1}{\mathcal{E}_{\pi }^2-\rho_\pi^2}       (-64)    \Big( 2 \mathcal{E}_{\pi } \rho_\pi^2 \left(\rho _{\tau }^2 \left(1-\rho_\pi^2\right)-3 \rho _{\tau }^4 - \rho_\pi^2\right) \nn \\
        && \hskip3truecm
+~2 \mathcal{E}_{\pi }^2 \rho_\pi^2 \left(1+2 \rho _{\tau }^2-\rho _{\tau }^4\right)
- 4 \mathcal{E}_{\pi }^3 \rho _{\tau }^2 \left(1-\rho _{\tau }^2\right) \nn\\
&& \hskip3truecm +~\rho _{\tau }^4 \rho_\pi^2 \left(1+3 \rho_\pi^2\right)-\rho_\pi^4+\rho_\pi^6 \Big) ~, \nn\\
S_{0T,2} &=& -4 S_{\bot T,2} = -4 S_{\parallel T,2} \nn\\
&=& \frac{1}{\mathcal{E}_{\pi }^2-\rho_\pi^2}      64 \Big( 2 \mathcal{E}_{\pi } \rho_\pi^2 \left(1+\rho _{\tau }^2\right) \left(\rho _{\tau }^2+3 \rho_\pi^2\right) \nn \\
        && \hskip2.5truecm
-~2 \mathcal{E}_{\pi }^2 \left(6 \rho _{\tau }^2 \rho_\pi^2+\rho _{\tau }^4 \left(2+\rho_\pi^2\right)+\rho_\pi^2\right)
+ 4 \mathcal{E}_{\pi }^3 \rho _{\tau }^2 \left(1+\rho _{\tau }^2\right) \nn\\
&& \hskip2.5truecm  +~\rho _{\tau }^4 \rho_\pi^2 \left(1-\rho_\pi^2\right)-\rho_\pi^4-3 \rho_\pi^6 \Big) ~, \nn\\
S_{\bot T,1} = S_{\parallel T,1} &=& \frac{ 1}{\mathcal{E}_{\pi }^2-\rho_\pi^2}    (-16) \Big( 2 \mathcal{E}_{\pi } \rho_\pi^2 \left(\rho _{\tau }^2 \left(3+\rho_\pi^2\right)-5 \rho _{\tau }^4+\rho_\pi^2\right) \nn \\
         && \hskip2.5truecm
-~\mathcal{E}_{\pi }^2 \left(4 \rho _{\tau }^2 \rho_\pi^2+\rho _{\tau }^4 \left(4+6 \rho_\pi^2\right)-2 \rho_\pi^2\right)
- 4 \mathcal{E}_{\pi }^3 \rho _{\tau }^2 \left(1-3 \rho _{\tau }^2\right) \nn\\
&& \hskip2.5truecm  +~\rho _{\tau }^4 \rho_\pi^2 \left(3+5 \rho_\pi^2\right)-3 \rho_\pi^4-\rho_\pi^6 \Big)  ~.
\eea

The expressions for the $R_i$ factors are
\bea
\label{Rifactors}
R_{t0} = 2\sqrt{2} R_{t \parallel} &=& \rho_\tau R_{SP0} = 2\sqrt{2} \rho_\tau R_{SP \parallel}
= \frac{  -32 \rho _{\tau }^2 \left(1-\mathcal{E}_{\pi }\right)
\left(2 \mathcal{E}_{\pi } \rho _{\tau }^2-\rho _{\tau }^4-\rho_\pi^2\right) }
{\sqrt{\mathcal{E}_{\pi }^2-\rho_\pi^2}} ~, \nn\\
R_{0\parallel } &=& \frac{ 2 \sqrt{2} \rho _{\tau }^2}{\mathcal{E}_{\pi }^2-\rho_\pi^2}
\Big( 2 \mathcal{E}_{\pi } \left(1+\rho _{\tau }^2\right) \left(3 \rho _{\tau }^2+\rho_\pi^2\right)-2 \mathcal{E}_{\pi }^2 \left(1+6 \rho _{\tau }^2+\rho _{\tau }^4+2 \rho_\pi^2\right) \nn \\
&& \hskip2truecm +~4 \mathcal{E}_{\pi }^3 \left(1+\rho _{\tau }^2\right)
- \rho _{\tau }^4 \left(3+\rho_\pi^2\right)-\rho_\pi^2+\rho_\pi^4 \Big) ~, \nn\\
R_{\parallel \bot } = -\sqrt{2} R_{0 \bot} &=& \frac{ 16 \rho _{\tau }^2 \left(1-2 \mathcal{E}_{\pi }+\rho_\pi^2\right) \left(\rho _{\tau }^2-\mathcal{E}_{\pi }\right)}{\sqrt{\mathcal{E}_{\pi }^2-\rho_\pi^2}}   ~, \nn\\
R_{SP t} &=& 32 \rho _{\tau } \left(+2 \mathcal{E}_{\pi } \rho _{\tau }^2-\rho _{\tau }^4-\rho_\pi^2\right) ~, \nn\\
R_{SP 0T} = 2\sqrt{2} R_{SP \parallel T} &=& \frac{1}{\rho_\tau} R_{0Tt} = \frac{2\sqrt{2}}{\rho_\tau} R_{\parallel T t}
= \frac{ 128 \left(\mathcal{E}_{\pi }-\rho_\pi^2\right) \left(2 \mathcal{E}_{\pi } \rho _{\tau }^2-\rho _{\tau }^4-\rho_\pi^2\right)}{\sqrt{\mathcal{E}_{\pi }^2-\rho_\pi^2}}   ~, \nn\\
&& \hskip-4truecm R_{0T 0,1} ~=~ \frac{ 32 \rho _{\tau } \Big( \mathcal{E}_{\pi } \left(1+\rho_\pi^2\right) \left(\rho _{\tau }^4+\rho_\pi^2\right)-4 \mathcal{E}_{\pi }^2 \rho _{\tau }^2 \left(1+\rho_\pi^2\right)+4 \mathcal{E}_{\pi }^3 \rho _{\tau }^2+2 \rho_\pi^2 \left(\rho _{\tau }^2-\rho_\pi^2\right) \left(1-\rho _{\tau }^2\right) \Big)}{\mathcal{E}_{\pi }^2-\rho_\pi^2}   ~, \nn\\
R_{0T0,2} = 2\sqrt{2} R_{0T \parallel} &=& \frac{ 32 \rho _{\tau }}{\mathcal{E}_{\pi }^2-\rho_\pi^2}     \Big(  \mathcal{E}_{\pi } \left( 8 \rho _{\tau }^2 \rho_\pi^2+3 \rho _{\tau }^4 \left(1+\rho_\pi^2\right)+3 \rho_\pi^2+3 \rho_\pi^4\right)-4 \mathcal{E}_{\pi }^2 \left(1+\rho _{\tau }^2\right) \left(\rho _{\tau }^2+\rho_\pi^2\right)  \nn \\
&& \hskip2truecm +~4 \mathcal{E}_{\pi }^3 \rho _{\tau }^2
-2 \rho_\pi^2 \left(1+\rho _{\tau }^2\right) \left(\rho _{\tau }^2+\rho_\pi^2\right) \Big)   ~, \nn\\
R_{0T\parallel T} &=& \frac{32 \sqrt{2}}{\mathcal{E}_{\pi }^2-\rho_\pi^2}       \Big( 2 \mathcal{E}_{\pi } \rho_\pi^2 \left(1+\rho _{\tau }^2\right) \left(3 \rho_\pi^2+\rho _{\tau }^2\right)-2 \mathcal{E}_{\pi }^2 \left(6 \rho _{\tau }^2 \rho_\pi^2+\rho _{\tau }^4 \left(2+\rho_\pi^2\right)+\rho_\pi^2\right) \nn \\
                && \hskip2truecm +~4 \mathcal{E}_{\pi }^3 \rho _{\tau }^2 \left(1+\rho _{\tau }^2\right)
+ \rho _{\tau }^4 \rho_\pi^2 \left(1-\rho_\pi^2\right)-\rho_\pi^4-3 \rho_\pi^6 \Big)   ~, \nn\\
R_{\bot T\parallel } &=& -\sqrt{2} R_{0T \bot} = -\sqrt{2} R_{\bot T 0} = R_{\parallel T \bot}
= \frac{ -32 \rho _{\tau } \left(1-2 \mathcal{E}_{\pi }+\rho_\pi^2\right) \left(\rho _{\tau }^4-\rho_\pi^2\right)}{\sqrt{\mathcal{E}_{\pi }^2-\rho_\pi^2}}   ~, \nn\\
R_{\bot T\parallel T} = -\sqrt{2} R_{0T \bot T}
&=& \frac{ 256 \rho _{\tau }^2 \left(1-2 \mathcal{E}_{\pi }+\rho_\pi^2\right) \left(\mathcal{E}_{\pi } \rho _{\tau }^2-\rho_\pi^2\right)}{\sqrt{\mathcal{E}_{\pi }^2-\rho_\pi^2}}  ~, \nn\\
R_{\parallel T0} &=& -\sqrt{2} R_{\bot T \bot, 2} = -\sqrt{2} R_{\parallel T \parallel, 2} \nn\\
&=& \frac{8 \sqrt{2} \rho_{\tau }}{\mathcal{E}_{\pi }^2-\rho_\pi^2}         \Big(  \mathcal{E}_{\pi } \left(8 \rho _{\tau }^2 \rho_\pi^2+3 \rho _{\tau }^4 \left(1+\rho_\pi^2\right)+3 \rho_\pi^2+3 \rho_\pi^4\right)-4 \mathcal{E}_{\pi }^2 \left(1+\rho _{\tau }^2\right) \left(\rho_\pi^2+\rho _{\tau }^2\right)  \nn \\
&& \hskip2truecm  +~4 \mathcal{E}_{\pi }^3 \rho _{\tau }^2
- 2 \rho_\pi^2 \left(1+\rho _{\tau }^2\right) \left(\rho_\pi^2+\rho _{\tau }^2\right) \Big)  ~, \nn\\
R_{\parallel T\parallel ,1} = R_{\bot T \bot, 1}
&=& \frac{-8 \rho_{\tau }}{\mathcal{E}_{\pi }^2-\rho_\pi^2}       \Big(  \mathcal{E}_{\pi } \left(8 \rho _{\tau }^2 \rho_\pi^2+\rho _{\tau }^4 \left(1+\rho_\pi^2\right)+\rho_\pi^2+\rho_\pi^4\right)-4 \mathcal{E}_{\pi }^2 \left(1-\rho _{\tau }^2\right) \left(\rho_\pi^2-\rho _{\tau }^2\right) \nn  \\
                      && \hskip2truecm -~4 \mathcal{E}_{\pi }^3 \rho _{\tau }^2
- 2 \rho_\pi^2 \left(3 \rho _{\tau }^2 \left(1+\rho_\pi^2\right)-\rho _{\tau }^4-\rho_\pi^2\right) \Big) ~.
\eea

And finally the expressions for the $I_i$ factors are
\bea
\label{Iifactors}
I_{t\bot } = \rho_\tau I_{SP\bot} &=& \frac{  8 \sqrt{2} \rho _{\tau }^2 \left(1-\mathcal{E}_{\pi }\right) \left(2 \mathcal{E}_{\pi } \rho _{\tau }^2-\rho _{\tau }^4-\rho_\pi^2\right) }{\sqrt{\mathcal{E}_{\pi }^2-\rho_\pi^2}}  ~, \nn\\
I_{\parallel \bot } = \sqrt{2}I_{0\bot} &=& \frac{-4 \rho _{\tau }^2}{\mathcal{E}_{\pi }^2-\rho_\pi^2}        \Big( 2 \mathcal{E}_{\pi } \left(1+\rho _{\tau }^2\right) \left(\rho_\pi^2+3 \rho _{\tau }^2\right)
- 2 \mathcal{E}_{\pi }^2 \left(1+6 \rho _{\tau }^2+\rho _{\tau }^4+2 \rho_\pi^2\right)
 \nn \\
&& \hskip2.5truecm
+~4 \mathcal{E}_{\pi }^3 \left(1+\rho _{\tau }^2\right)
-\left(3+\rho_\pi^2\right) \rho _{\tau }^4-\rho_\pi^2+\rho_\pi^4 \Big)  ~, \nn\\
I_{0T\parallel } = - I_{\parallel T 0}
&=& \frac{ -16 \sqrt{2} \rho _{\tau } \left(1-2 \mathcal{E}_{\pi }+\rho_\pi^2\right) \left(\rho _{\tau }^4-\rho_\pi^2\right)}{\sqrt{\mathcal{E}_{\pi }^2-\rho_\pi^2}}   ~, \nn\\
I_{\bot Tt} = -\rho_\tau I_{SP \bot T} &=& \frac{ 32 \sqrt{2} \rho _{\tau } \left(\mathcal{E}_{\pi }-\rho_\pi^2\right) \left(2 \mathcal{E}_{\pi } \rho _{\tau }^2-\rho _{\tau }^4-\rho_\pi^2\right)}{\sqrt{\mathcal{E}_{\pi }^2-\rho_\pi^2}}      ~, \nn\\
I_{\bot T0} &=& -I_{0T \bot} = \frac{1}{\sqrt{2}} I_{\bot T \parallel} = - \frac{1}{\sqrt{2}} I_{\parallel T \bot} \nn\\
&=& \frac{8 \sqrt{2} \rho _{\tau }}{\mathcal{E}_{\pi }^2-\rho_\pi^2}    \Big(  \mathcal{E}_{\pi } \left(8 \rho_\pi^2 \rho _{\tau }^2+3 \rho _{\tau }^4 \left(1+\rho_\pi^2\right)+3 \rho_\pi^2+3 \rho_\pi^4\right) \nn \\
&& ~~~~~
-~4 \mathcal{E}_{\pi }^2 \left(1+\rho _{\tau }^2\right) \left(\rho_\pi^2+\rho _{\tau }^2\right)+4 \mathcal{E}_{\pi }^3 \rho_{\tau }^2
- 2 \rho_\pi^2 \left(1+\rho _{\tau }^2\right) \left(\rho _{\tau }^2+\rho_\pi^2\right) \Big)   ~, \nn\\
I_{\parallel T\bot } = \sqrt{2}I_{0T \bot}
&=& \frac{-16 \rho _{\tau }}{\mathcal{E}_{\pi }^2-\rho_\pi^2}       \Big(   \mathcal{E}_{\pi } \left(8 \rho _{\tau }^2 \rho_\pi^2+3 \rho _{\tau }^4 \left(1+\rho_\pi^2\right)+3 \rho_\pi^2+3 \rho_\pi^4\right) \nn \\
&& \hskip2.5truecm
-~4 \mathcal{E}_{\pi }^2 \left(1+\rho _{\tau }^2\right) \left(\rho_\pi^2+\rho _{\tau }^2\right)+4 \mathcal{E}_{\pi }^3 \rho _{\tau }^2  \nn\\
&& \hskip2.5truecm
-~2 \rho_\pi^2 \left(1+\rho _{\tau }^2\right) \left(\rho _{\tau }^2+\rho_\pi^2\right)  \Big)  ~.
\eea

\bibliographystyle{JHEP}
\bibliography{ref}

\end{document}